\input harvmac
\let\includefigures=\iftrue
\let\useblackboard=\iftrue
\newfam\black

\includefigures
\message{If you do not have epsf.tex (to include figures),}
\message{change the option at the top of the tex file.}
\input epsf
\def\figin{\epsfcheck\figin}\def\figins{\epsfcheck\figins}
\def\epsfcheck{\ifx\epsfbox\UnDeFiNeD
\message{(NO epsf.tex, FIGURES WILL BE IGNORED)}
\gdef\figin##1{\vskip2in}\gdef\figins##1{\hskip.5in}
\else\message{(FIGURES WILL BE INCLUDED)}%
\gdef\figin##1{##1}\gdef\figins##1{##1}\fi}
\def\DefWarn#1{}
\def\figinsert{\goodbreak\midinsert}
\def\ifig#1#2#3{\DefWarn#1\xdef#1{fig.~\the\figno}
\writedef{#1\leftbracket fig.\noexpand~\the\figno}%
\figinsert\figin{\centerline{#3}}\medskip\centerline{\vbox{
\baselineskip12pt\advance\hsize by -1truein
\noindent\footnotefont{\bf Fig.~\the\figno:} #2}}
\endinsert\global\advance\figno by1}
\else
\def\ifig#1#2#3{\xdef#1{fig.~\the\figno}
\writedef{#1\leftbracket fig.\noexpand~\the\figno}%
\global\advance\figno by1} \fi

\def\id{{1 \kern-.28em {\rm l}}}

\def\K3{{\bf K3}}
\def\journal#1&#2(#3){\unskip, \sl #1\ \bf #2 \rm(19#3) }
\def\andjournal#1&#2(#3){\sl #1~\bf #2 \rm (19#3) }

\def\bar{\overline}
\def\hat{\widehat}
\def\ie{{\it i.e.}}
\def\eg{{\it e.g.}}

\def\tilde{\widetilde}

\def\frac#1#2{{#1\over#2}}

\def\half{\frac12}

\def\d{\partial}

\def\inbar{\,\vrule height1.5ex width.4pt depth0pt}
\def\IC{\relax\hbox{$\inbar\kern-.3em{\rm C}$}}
\def\IR{\relax{\rm I\kern-.18em R}}
\def\IP{\relax{\rm I\kern-.18em P}}

%
%

%
\catcode`\@=11
\def\slash#1{\mathord{\mathpalette\c@ncel{#1}}}
\overfullrule=0pt

\def\lam{\lambda}

\def\underrel#1\over#2{\mathrel{\mathop{\kern\z@#1}\limits_{#2}}}

\catcode`\@=12


%

\def\det{{\rm det}}

\def \sinh{{\rm sinh}}
\def \cosh{{\rm cosh}}

\def\det{{\rm det}}
\def\exp{{\rm exp}}


\lref\KutasovCT{
  D.~Kutasov,
  ``A geometric interpretation of the open string tachyon,''
  arXiv:hep-th/0408073.
}

\lref\KutasovDJ{
  D.~Kutasov,
  ``D-brane dynamics near NS5-branes,''
  arXiv:hep-th/0405058.
}

\lref\SenNF{
  A.~Sen,
  ``Tachyon dynamics in open string theory,''
  Int.\ J.\ Mod.\ Phys.\  A {\bf 20}, 5513 (2005)
  [arXiv:hep-th/0410103].
}

\lref\SenTM{
  A.~Sen,
  ``Dirac-Born-Infeld action on the tachyon kink and
    vortex,''
  Phys.\ Rev.\  D {\bf 68}, 066008 (2003)
  [arXiv:hep-th/0303057].
}

\lref\SenCZ{
  A.~Sen,
  ``Geometric tachyon to universal open string tachyon,''
  JHEP {\bf 0705}, 035 (2007)
  [arXiv:hep-th/0703157].
}

\lref\KutasovER{
  D.~Kutasov and V.~Niarchos,
  ``Tachyon effective actions in open string theory,''
  Nucl.\ Phys.\  B {\bf 666}, 56 (2003)
  [arXiv:hep-th/0304045].
}

\lref\TseytlinDJ{
  A.~A.~Tseytlin,
  ``Born-Infeld action, supersymmetry and string theory,''
  arXiv:hep-th/9908105.
}

\lref\LambertZR{
  N.~D.~Lambert, H.~Liu and J.~M.~Maldacena,
  ``Closed strings from decaying D-branes,''
  JHEP {\bf 0703}, 014 (2007)
  [arXiv:hep-th/0303139].
}

\lref\CallanAT{
C.~G.~Callan, J.~A.~Harvey and A.~Strominger,
``Supersymmetric string solitons,''
arXiv:hep-th/9112030.
}

\lref\AharonyUB{
O.~Aharony, M.~Berkooz, D.~Kutasov and N.~Seiberg,
``Linear dilatons, NS5-branes and holography,''
JHEP {\bf 9810}, 004 (1998)
[arXiv:hep-th/9808149].
}

\lref\GiveonPX{
A.~Giveon and D.~Kutasov,
``Little string theory in a double scaling limit,''
JHEP {\bf 9910}, 034 (1999)
[arXiv:hep-th/9909110].
}

\lref\ItzhakiDD{ N.~Itzhaki, J.~M.~Maldacena, J.~Sonnenschein and
S.~Yankielowicz, ``Supergravity and the large N limit of theories
with sixteen  supercharges,'' Phys.\ Rev.\ D {\bf 58}, 046004
(1998) [arXiv:hep-th/9802042].
}

\lref\agan{
M.~Aganagic, C.~Popescu and J.~H.~Schwarz,
  ``D-brane actions with local kappa symmetry,''
  Phys.\ Lett.\  B {\bf 393}, 311 (1997)
  [arXiv:hep-th/9610249];
  ``Gauge-invariant and gauge-fixed D-brane actions,''
  Nucl.\ Phys.\  B {\bf 495}, 99 (1997)
  [arXiv:hep-th/9612080].
}

\lref\Refkappaone{
E.~Bergshoeff and P.~K.~Townsend, ``Super D-branes,''
  Nucl.\ Phys.\  B {\bf 490}, 145 (1997)
  [arXiv:hep-th/9611173].
}

\lref\refA{}

\lref\CalMal{
C.~G.~Callan and J.~M.~Maldacena,
  Nucl.\ Phys.\  B {\bf 513}, 198 (1998)
  [arXiv:hep-th/9708147].
}

\lref\MaldRey{
S.~J.~Rey and J.~T.~Yee,
  Eur.\ Phys.\ J.\  C {\bf 22}, 379 (2001)
  [arXiv:hep-th/9803001];

J.~M.~Maldacena,
  Phys.\ Rev.\ Lett.\  {\bf 80}, 4859 (1998)
  [arXiv:hep-th/9803002].
}

\lref\GDO{
N.~Drukker, D.~J.~Gross and H.~Ooguri,
  Phys.\ Rev.\  D {\bf 60}, 125006 (1999)
  [arXiv:hep-th/9904191].
}

\lref\CalGui{
C.~G.~Callan and A.~Guijosa,
  Nucl.\ Phys.\  B {\bf 565}, 157 (2000)
  [arXiv:hep-th/9906153].
}

\lref\HKKKY{
C.~P.~Herzog, A.~Karch, P.~Kovtun, C.~Kozcaz and L.~G.~Yaffe,
  JHEP {\bf 0607}, 013 (2006)
  [arXiv:hep-th/0605158].
}

\lref\Drag{
S.~S.~Gubser,
  Phys.\ Rev.\  D {\bf 74}, 126005 (2006)
  [arXiv:hep-th/0605182].
}

\lref\NakayamaYX{
  Y.~Nakayama, Y.~Sugawara and H.~Takayanagi,
  ``Boundary states for the rolling D-branes in NS5 background,''
  JHEP {\bf 0407}, 020 (2004)
  [arXiv:hep-th/0406173].
}

\lref\SahakyanCQ{
  D.~A.~Sahakyan,
  ``Comments on D-brane dynamics near NS5-branes,''
  JHEP {\bf 0410}, 008 (2004)
  [arXiv:hep-th/0408070].
}

\lref\NakayamaGE{
  Y.~Nakayama, K.~L.~Panigrahi, S.~J.~Rey and H.~Takayanagi,
  ``Rolling down the throat in NS5-brane background: The case of  electrified
  D-brane,''
  JHEP {\bf 0501}, 052 (2005)
  [arXiv:hep-th/0412038].
}

\lref\KachruSX{
  S.~Kachru, R.~Kallosh, A.~Linde, J.~M.~Maldacena, L.~P.~McAllister and S.~P.~Trivedi,
  ``Towards inflation in string theory,''
  JCAP {\bf 0310}, 013 (2003)
  [arXiv:hep-th/0308055].
}

\lref\SakaiCN{
  T.~Sakai and S.~Sugimoto,
  ``Low energy hadron physics in holographic QCD,''
  Prog.\ Theor.\ Phys.\  {\bf 113}, 843 (2005)
  [arXiv:hep-th/0412141].
}

\lref\AntonyanVW{
  E.~Antonyan, J.~A.~Harvey, S.~Jensen and D.~Kutasov,
  ``NJL and QCD from string theory,''
  arXiv:hep-th/0604017.
}

\lref\OoguriBG{
  H.~Ooguri and Y.~Ookouchi,
  ``Meta-stable supersymmetry breaking vacua on intersecting branes,''
  Phys.\ Lett.\  B {\bf 641}, 323 (2006)
  [arXiv:hep-th/0607183].
}

\lref\FrancoHT{
  S.~Franco, I.~Garcia-Etxebarria and A.~M.~Uranga,
  ``Non-supersymmetric meta-stable vacua from brane configurations,''
  JHEP {\bf 0701}, 085 (2007)
  [arXiv:hep-th/0607218].
}

\lref\BenaRG{
  I.~Bena, E.~Gorbatov, S.~Hellerman, N.~Seiberg and D.~Shih,
  ``A note on (meta)stable brane configurations in MQCD,''
  JHEP {\bf 0611}, 088 (2006)
  [arXiv:hep-th/0608157].
}

\lref\GiveonFK{
  A.~Giveon and D.~Kutasov,
  ``Gauge symmetry and supersymmetry breaking from intersecting branes,''
  Nucl.\ Phys.\  B {\bf 778}, 129 (2007)
  [arXiv:hep-th/0703135].
}

\lref\GiveonEW{
  A.~Giveon and D.~Kutasov,
  ``Stable and Metastable Vacua in Brane Constructions of SQCD,''
  JHEP {\bf 0802}, 038 (2008)
  [arXiv:0710.1833 [hep-th]].
}

\lref\HarveyNA{
  J.~A.~Harvey, D.~Kutasov and E.~J.~Martinec,
  ``On the relevance of tachyons,''
  arXiv:hep-th/0003101.
}

\lref\FelderSV{
  G.~N.~Felder, L.~Kofman and A.~Starobinsky,
  ``Caustics in tachyon matter and other Born-Infeld scalars,''
  JHEP {\bf 0209}, 026 (2002)
  [arXiv:hep-th/0208019].
}

\lref\FelderXU{
  G.~N.~Felder and L.~Kofman,
  ``Inhomogeneous fragmentation of the rolling tachyon,''
  Phys.\ Rev.\  D {\bf 70}, 046004 (2004)
  [arXiv:hep-th/0403073].
}

\lref\LukyanovNJ{
  S.~L.~Lukyanov, E.~S.~Vitchev and A.~B.~Zamolodchikov,
  ``Integrable model of boundary interaction: The paperclip,''
  Nucl.\ Phys.\  B {\bf 683}, 423 (2004)
  [arXiv:hep-th/0312168].
}

\lref\HananyIE{
  A.~Hanany and E.~Witten,
  ``Type IIB superstrings, BPS monopoles, and three-dimensional gauge
  dynamics,''
  Nucl.\ Phys.\  B {\bf 492}, 152 (1997)
  [arXiv:hep-th/9611230].
}

\lref\PeskinEV{
  M.~E.~Peskin and D.~V.~Schroeder,
  ``An Introduction To Quantum Field Theory,''
{\it  Reading, USA: Addison-Wesley (1995) 842 p}, Section 20.1, p. 690-691.
}

\lref\PolchinskiRR{
  J.~Polchinski,
  ``String theory. Vol. 2: Superstring theory and beyond,''
{\it  Cambridge, UK: Univ. Pr. (1998) 531 p}.
}

\lref\HellermanZM{
  S.~Hellerman,
  ``On the landscape of superstring theory in $D > 10$,''
  arXiv:hep-th/0405041.
}

\lref\AharonyAN{
  O.~Aharony and D.~Kutasov,
  ``Holographic Duals of Long Open Strings,''
  Phys.\ Rev.\  D {\bf 78}, 026005 (2008)
  [arXiv:0803.3547 [hep-th]].
}

\lref\CallanUB{
  C.~G.~Callan, I.~R.~Klebanov, A.~W.~W.~Ludwig and J.~M.~Maldacena,
  ``Exact solution of a boundary conformal field theory,''
  Nucl.\ Phys.\  B {\bf 422}, 417 (1994)
  [arXiv:hep-th/9402113].
}

\lref\McNeesKM{
  R.~McNees, R.~C.~Myers and A.~Sinha,
  ``On quark masses in holographic QCD,''
  JHEP {\bf 0811}, 056 (2008)
  [arXiv:0807.5127 [hep-th]].
}

\Title{}
{\vbox{\centerline{Brane -- Antibrane Dynamics}
\bigskip
\centerline{From the Tachyon DBI Action}
}}
\bigskip

\centerline{\it Denis Erkal, David Kutasov  and
Oleg Lunin}
\bigskip

\centerline{EFI and Department of Physics, University of
Chicago}\centerline{5640 S. Ellis Av. Chicago, IL 60637}

\smallskip

\vglue .3cm

\bigskip

\bigskip
\noindent
The Tachyon-Dirac-Born-Infeld (TDBI) action captures some
aspects of the dynamics of non-BPS $D$-branes in type II string
theory. We show that it can also be used to study the classical
interactions of BPS branes and antibranes. Our analysis sheds
light on real time $D-\bar D$ tachyon condensation, on the
proposal that the tachyon field can be thought of as an extra
spatial dimension whose role is similar to the radial direction
in holography, and on A. Sen's open string completeness conjecture.

\bigskip

\Date{January 2009}

\newsec{Introduction}

Non-supersymmetric excitations of supersymmetric vacua in weakly
coupled string theory can be (roughly) divided into three classes:
\item{(1)} States whose energy remains finite as the string coupling
$g_s\to 0$ correspond to fundamental string excitations. If
the energy is of order one in string units, they can be described in
terms of vertex operators. When the energy in string units is large,
the typical states correspond to highly excited and/or long strings
and other descriptions may be more useful.
\item{(2)} States with energy of order $1/g_s$ correspond to
non-supersymmetric $D$-branes. Examples include non-BPS $D$-branes
(see \eg\ \SenNF\ for a review) and classical deformations of BPS
$D$-branes of the sort described in section 4 below. Such states can be
studied to leading order in $g_s$ by using classical open string theory.
Gravitational back reaction is subleading in $g_s$, as for the states
in point (1) above.
\item{(3)} States with energy of order $1/g_s^2$ correspond to deformations
of the gravitational background. To study them one has to analyze
the corresponding worldsheet CFT on the sphere. Gravitational backreaction
is a leading order effect and cannot be neglected. Examples of such states
are non-extremal Neveu-Schwarz (NS) fivebranes.
\smallskip
\noindent
In this paper we will discuss the dynamics of certain non-supersymmetric
$D$-brane systems. Since these systems are in general unstable, their study
is intimately tied to the problem of time dependence in string theory,
which is of much interest, \eg\ in the context of early universe cosmology.
Two examples of such systems in which the time
dependent dynamics has been analyzed in some detail are:
\item{(1)} The rolling tachyon solution (reviewed in \SenNF), which
corresponds to displacing the tachyon on a non-BPS $D$-brane from the maximum
of its potential, and studying its subsequent time evolution.
\item{(2)} A $D$-brane moving in the gravitational potential of a stack of
$NS5$-branes \KutasovDJ\ (see also \refs{\NakayamaYX\SahakyanCQ-\NakayamaGE}).
In this case, the role of the tachyon is played by the distance of the $D$-brane
from the $NS5$-branes.

\noindent
In both of these cases the relevant time dependent solution is a Wick rotated
version of a solvable worldsheet CFT in a Euclidean signature target space.
One can use this CFT to calculate physical observables, such as the stress
tensor and the amplitude for closed string production, to all orders in $\alpha'$.

An interesting outcome of the study of these solutions is the observation that
some of the exact results obtained from the worldsheet analysis are captured
by an effective action of the Dirac-Born-Infeld (DBI) type, which includes the
tachyon  \SenNF. This action has the form\foot{Here and below we set $\alpha'=1$.}
\eqn\effact{S=-\int d^{p+1}x V(T)\sqrt{-\det A}~,}
where ($\mu,\nu=0,1,\cdots, p$)
\eqn\defaa{A_{\mu\nu}=\eta_{\mu\nu}+\partial_\mu T\partial_\nu T+
\partial_\mu y^I\partial_\nu y^I+2\pi F_{\mu\nu}~,}
$T$ is the tachyon field, the scalar fields $y^I$, $I=p+1,\cdots, 9$, parameterize the
location of the $D$-brane in the transverse directions, $F_{\mu\nu}$ is the field strength
of the worldvolume gauge field, and we have omitted the worldvolume fermions
and couplings to massless closed string fields (we will comment on them in
section 3). The tachyon potential $V(T)$ is given by
\eqn\vvv{V(T)={\tau_p\over\cosh{T\over\sqrt2}}}
for the non-BPS $D$-brane, whose tension is $\tau_p$, and by
\eqn\vdns{V(T)=2\tau_pe^{-{T\over\sqrt k}}}
for the case of a $D$-brane falling onto a stack of $k$ $NS5$-branes.
The coefficient of the exponential in \vdns\ can be modified by shifting $T$,
but as we will see below, the choice in \vdns\ is natural.

For the non-BPS $D$-brane, the existence of an effective action that includes
the tachyon is at first sight surprising, since the tachyon has a string scale
(negative) mass squared. It was shown in \KutasovER\ that
this action can be thought of as describing small fluctuations about the rolling
tachyon background (or, more precisely, about the so called ``half S-brane'').
This background plays a role similar to that of the solutions with constant brane
electromagnetic field and/or velocity that lead to the usual DBI action (see
\eg\ \TseytlinDJ\ for a review).

For the $D/NS$ system, \effact\ is the usual DBI action (since,
as mentioned above, the tachyon is a geometric mode).
It is expected to provide a good description of $D$-brane propagation in
the vicinity of $k$ $NS5$-branes\foot{At least as long as the local string
coupling remains small.} for large $k$. However, it turns out to be
useful for $k$ of order one as well \KutasovDJ, for reasons similar to
those described in the previous paragraph.

Unlike \vvv, the potential \vdns\ does not have a local maximum (which in the
case \vvv\ corresponds to the non-BPS $D$-brane). To get a system that more closely
resembles the non-BPS $D$-brane, it is useful to compactify one of the directions
transverse to the $NS5$-branes on a circle \KutasovCT. This leads to the
appearance of a local maximum of the energy functional, corresponding to a
$D$-brane localized across the circle from the $NS5$-branes. The potential \vdns\
is modified to
\eqn\vdnscircle{V(T)={\tau_p\over\cosh{T\over\sqrt k}}~,}
where $T$ parameterizes the location of the $Dp$-brane on the circle. $T=0$
corresponds to the local maximum (the antipodal configuration); $\tau_p$
is the tension of the $D$-brane in that configuration. $T\to\pm \infty$
corresponds to the $D$-brane approaching the fivebranes from either side.
The potential \vdnscircle\ reduces to \vdns\ as $T\to\infty$.

Interestingly, the potential \vdnscircle\ is very similar, and for $k=2$
(two $NS5$-branes) identical, to that of the non-BPS brane system \vvv. We will
utilize this analogy below, by using the former to gain insight
into the latter. The similarity of \vvv\ and \vdnscircle\ as well as other
considerations \refs{\KutasovCT,\SenCZ} hint at a deeper relation between the
two systems, but we will not explore this relation further here.

The question that motivated this investigation is whether one can extend the
understanding of the dynamics of the tachyon on non-BPS $D$-branes provided by
the action \effact\ -- \vdnscircle\ to the $D-\bar D$ system, which has a
(complex) tachyon in the $D-\bar D$ open string sector. This system plays an
important role in string theory realizations of inflation (see \eg\ \KachruSX),
holographic QCD \refs{\SakaiCN,\AntonyanVW}, and supersymmetry breaking brane
configurations \refs{\OoguriBG\FrancoHT\BenaRG\GiveonFK-\GiveonEW}.
In all these applications it would be nice to have a better understanding of the
dynamics of the tachyon.

One may try to write actions analogous to \effact, consisting of DBI actions
for the brane and antibrane, coupled via the tachyon, but such actions are not
well understood, and this approach has not met with much success. We will pursue
a different route -- start from the well motivated action \effact\ for the non-BPS
$D$-brane, and use the fact that it describes BPS branes and antibranes as kinks
and anti-kinks of the tachyon field. We will see that this approach leads to a useful
description of brane-antibrane dynamics, and in particular incorporates the $D-\bar D$
tachyon.

For $D$-branes propagating in the vicinity of $k$ $NS5$-branes the tachyon DBI action \effact\ 
provides a reliable description of the $D-\bar D$ system at large $k$. For the brane-antibrane 
system in critical string theory and for the $D/NS$ system with $k$ of order one it is expected to
provide a qualitative guide to the dynamics; the experience from the non-BPS brane case
leads one to expect that in some cases it may be quantitatively accurate as well.

One can also view the TDBI action as defining a (string inspired) model that exhibits in a
field theoretic context many of the non-trivial features normally associated with classical 
open string theory.  As such, it may be useful for constructing models of hybrid inflation and 
other applications.

The plan of the rest of the paper is as follows. In section 2 we discuss some background
material. We review the rolling tachyon solution \SenNF\ of the TDBI action
\effact\ -- \vvv, focusing on the stress tensor of the brane, closed string production,
and A. Sen's proposal that this open string description is equivalent to the closed string
picture at leading order in $g_s$. We then review the $D/NS$ system,
pointing out some of the similarities between the two systems.

In section 3 we turn to a description of the BPS $D$-brane in terms of the TDBI action \effact.
The BPS brane is a singular
(\ie\ zero width) kink of the tachyon field, and is usually defined as a limit from non-singular
configurations that have non-zero width but do not solve the equations of motion (see \eg\
\SenTM). We show that one can simplify this description by thinking about the tachyon
direction as an extra dimension of space \KutasovCT, and viewing the BPS $D$-brane as one that is
extended in this direction. As a simple application, we show that the low energy dynamics on
the kink gives rise to the correct action for a BPS $D$-brane.

The above description of the BPS $D$-brane leads to a natural question: what is the dynamics of
worldvolume fields that depend on the tachyon, and what is their interpretation in the usual
$9+1$ dimensional spacetime? This is studied in section 4. We show that such excitations do
not give rise to new normalizable states on the $D$-brane. Instead they can be interpreted from
the $9+1$ dimensional point of view as describing closed string excitations with energy of order
$1/g_s$ that couple to the $D$-brane. The equations of motion of the TDBI action can be thought of
as describing the time evolution of these closed strings, after they are created in the vicinity
of the $D$-brane. We point out that the resulting picture is in agreement with A. Sen's open string
completeness proposal.

In section 5 we move on to a discussion of the $D-\bar D$ system. We show that the results
of sections  3, 4 lead to a nice picture of the classical dynamics of the $D-\bar D$ tachyon.
From the higher dimensional point of view, this tachyon can be thought of as localized at  large $T$.
Its condensation has a simple geometric interpretation, which provides further support for the
open string completeness proposal.

In section 6 we discuss a related system, which shares some properties with the ones we study.
The system involves $N$ $D3$-branes at large 't Hooft coupling, with a $D$-string ending on them. 
The $D3$-branes can be replaced by their near-horizon geometry, $AdS_5\times S^5$. The string 
is stretched in the radial direction of $AdS_5$ and can be studied in a way similar to the one used 
for the other systems. In this case the boundary theory is a standard gauge theory, and some of the 
things we discuss for the other systems can be understood using the  AdS/CFT correspondence.

We conclude and discuss our results in section 7. Some technical details appear in the appendices.

\newsec{Tachyon dynamics on non-BPS $D$-branes}

In this section we review some results on time dependent processes involving non-BPS $D$-branes,
focusing on aspects that will play a role in our subsequent discussion of non-BPS excitations of
BPS $D$-branes and $D-\bar D$ systems. In the first subsection we discuss non-BPS branes in type
II string theory on $\IR^{9,1}$  \SenNF. In the second we turn to $D$-branes propagating in the vicinity
of $NS5$-branes \KutasovDJ.

\subsec{Rolling tachyon}

Since the only field that is excited in these solutions is the tachyon, the Lagrangian \effact\
simplifies:
\eqn\simptach{S=-\int d^{p+1}x V(T)\sqrt{1+\partial_\mu T\partial^\mu T}~,}
with
\eqn\genv{V(T)={\tau_p\over\cosh\alpha T}~,}
where $\alpha=1/\sqrt2$ for the non-BPS $D$-brane, and $\alpha=1/\sqrt k$ for the $D/NS$ system.

The equation of motion of \simptach\ has a solution of the form
\eqn\rolling{\sinh\alpha T=C\cosh\alpha t~.}
This solution describes the tachyon climbing up the potential \genv,
reaching the minimal value
\eqn\mintt{\sinh\alpha T_0=C}
at $t=0$, and then rolling back to infinity at late times. The energy density of the solution
is given by
\eqn\ttoo{T_{00}={\tau_p\over\cosh\alpha T}\left(1-\dot T^2\right)^{-{1\over2}}
={\tau_p\over\sqrt{1+C^2}}<\tau_p~.}
Note that it is lower than that of the non-BPS $D$-brane, as expected.

The remaining non-vanishing components of the stress tensor are ($i,j=1,2,\cdots, p$
run over the spatial coordinates along the $D$-brane)
\eqn\ttij{T_{ij}=-\delta_{ij}{\tau_p\over\cosh\alpha T}\sqrt{1-\dot T^2}=
-\delta_{ij}{\tau_p\sqrt{1+C^2}\over 1+C^2\cosh^2\alpha t}=P(t)\delta_{ij}~.}
We see that the pressure $P$ goes exponentially to zero at late times, in agreement
with the results of the full classical open string analysis \SenNF.

As discussed in \KutasovCT, the form of the effective action \effact, \defaa\
suggests that the tachyon direction can be thought of as an additional spatial
dimension in which the non-BPS $D$-brane is localized. From this point of view,
\rolling\ describes the position of the brane in $T$ as a function of time. The
potential \genv\ gives rise to a force that pulls the brane towards large $|T|$. The
motion of the brane under the influence of this force is analogous to projectile
motion in a gravitational potential -- the brane comes in from large $T$, comes
to rest at $T_0$ \mintt\ at $t=0$, and then accelerates back out to $T\to\infty$.
Its speed in the $T$ direction approaches that of light at large  $t$.

The energy of the brane \ttoo\ is determined by its position at the turning point
of the trajectory \mintt. Thus, we can think of the tachyon direction as an analog
of the radial direction in holographic backgrounds -- position in this direction is
related to the energy of the process. Later we will see other manifestations of
this relation.

The rolling $D$-brane \rolling\ is a time dependent source for closed strings,
and one can calculate the amplitude for the production of on-shell closed strings
in this background. Consider, for example, the case of a non-BPS $D0$-brane in
type IIB string theory. It can emit all left-right symmetric closed string modes, with
arbitrary excitation level $N=N_L=N_R$, and transverse momentum $\vec k$.

For light closed string modes, such as the graviton and dilaton, one can calculate the
emission amplitude by coupling the TDBI action \effact\ to these fields and solving their
equations of motion in the time dependent background \rolling. For modes with arbitrary
excitation level $N$, one needs to perform the full worldsheet analysis. One finds
\LambertZR\ that the energy emitted into a given mode is finite, and behaves at large $N$ like
\eqn\emmamp{I(N)\sim e^{-2\pi\sqrt{2N}}~.}
Since the high energy density of left-right symmetric closed string states with arbitrary
spatial momentum $\vec k$ goes like
\eqn\dens{\rho(N)\sim N^{-\half}e^{2\pi\sqrt{2N}}~,}
the total energy emitted into modes with mass smaller than some fixed large value $M$ behaves like
\eqn\largem{\sum_{N\le M^2/4} N^{-\half}\sim M~.}
One can similarly compute the typical momentum of the emitted closed strings, and find
that it goes like $|\vec k|\sim \sqrt{M}$. Thus, their typical velocity is
$|\vec v|\sim M^{-\half}$, so for large $M$ they move away from the $D$-brane very slowly.

While the energy emitted into closed strings is suppressed by a factor of $g_s$ (since the energy
of the non-BPS $D$-brane is of order $1/g_s$ while that of the emitted closed strings scales like $g_s^0$),
it grows with the mass of the closed strings (see \largem). The calculations leading to \largem\ are
only valid for $M$ that remains finite in the limit $g_s\to 0$, but it is tempting to continue
the results to energies of order $1/g_s$. One is led to a picture where the non-BPS
$D0$-brane decays into massive closed strings with mass $\tau_0\sim 1/g_s$ and small
velocities.

At first sight it may seem that the production of massive closed strings invalidates the open string
picture described by the rolling tachyon solution, but this is not the case. In fact, the closed strings
with mass of order $1/g_s$ mentioned above {\it are} the $D0$-brane in the rolling tachyon state.
To leading order in $g_s$, their time evolution is described by classical open string theory, and is
well approximated by the TDBI action. The production of light closed strings, \ie\ those with finite
oscillator level $N$, can be calculated using the techniques of \LambertZR. It modifies the dynamics at
a subleading order in $g_s$.

In particular, one can interpret the stress-tensor \ttoo, \ttij\ as describing the heavy closed strings that
make up the $D0$-brane. From this perspective, the fact that the pressure goes to zero at late times is due to
their non-relativistic nature. This picture, known as the open string completeness conjecture, is due to A.
Sen, and is further discussed in \SenNF\ and references therein. Note that it agrees with the classification
of states mentioned in the introduction.

We also note that the above discussion shows that despite the fact that the non-BPS $D0$-brane has a string
scale tachyon on its worldvolume, its lifetime is much longer than $l_s$. The $D$-brane with condensing
tachyon can be replaced by a well localized massive closed string state. This state expands slowly
away from the location of the original $D0$-brane; the timescale associated with this
expansion diverges in the limit $g_s\to 0$.

\subsec{$D$-brane propagation in the vicinity of $NS5$-branes}

As mentioned above, the action \simptach, \genv\ also describes the motion of a $D$-brane in the vicinity
of a stack of $k$ $NS5$-branes, with one of the directions transverse to the fivebranes compactified on a
small circle \refs{\KutasovDJ,\KutasovCT}. The tachyon direction is now geometric, parameterizing
position on the circle. It also has a holographic interpretation in the context of Little String Theory (LST)
\AharonyUB.

From the bulk point of view, the ``rolling tachyon'' solution \rolling\ has in this case a simple
interpretation. It describes a $D0$-brane (say) thrown away from the fivebranes along the circle.
The $D0$-brane moves up to a maximal height \mintt\ and then falls back onto the fivebranes due to their
gravitational pull.

The stress tensor corresponding to this solution is given again by \ttoo, \ttij, and has the property
that the pressure goes to zero at late times. One can attempt to give it a similar interpretation to
the one discussed above for the case of non-BPS $D$-branes in flat space. To this end one needs to compute
the energy emission rate.

In this case, it is convenient to restrict the discussion to the region of large
$T$, where one can ignore the compactness of the circle, the potential $V(T)$ takes
the form \vdns, and the solution \rolling\ is
\eqn\roldns{\alpha T=\ln\cosh\alpha t+{\rm const}~.}
The reason is that this gives rise to a solvable worldsheet boundary CFT \refs{\KutasovDJ,\NakayamaYX},
which can be analyzed for all $k$.  The rate of emission into a given perturbative string state
at oscillator level $N$ is given by \refs{\NakayamaYX,\SahakyanCQ} (compare to \emmamp)
\eqn\emmampdns{I(N)\sim e^{-2\pi\sqrt{{2k-1\over k}N}}~.}
The density of left-right symmetric perturbative string states is
\eqn\densdns{\rho(N)\sim N^{-\half}e^{2\pi\sqrt{{2k-1\over k}N}}~.}
Combining the two leads again to \largem. Thus, we conclude as there that the typical
states emitted by the rolling $D$-brane are very heavy. The velocities of these closed
strings transverse to the $D$-brane but along the fivebranes are again very small. The
tachyon DBI action describes the properties of these strings, \eg\ their stress
tensor \ttoo, \ttij\ to leading order in $g_s$.

So far we discussed the dynamics from the point of view of the bulk $9+1$ dimensional
spacetime. To compare to the discussion of non-BPS $D$-branes in the previous subsection,
one has to take a boundary perspective, of an observer that lives on the fivebranes. This
is the analog to the perspective of  an observer that lives in $\IR^{9,1}$ in the previous case.

From that point of view, the direction along the circle is non-geometrical, rather like
the tachyon direction on a non-BPS $D$-brane, and the $D0$-brane moving in this direction
is localized in the $5+1$ dimensional spacetime. The theory on the fivebranes is in this
case the LST of $k$ $NS5$-branes (in type IIA string theory) and
the $D0$-brane is a non-BPS excitation in this theory.

Since much about the fivebrane theory is still not understood, it is difficult to perform
the analogs of the calculations of the previous subsection for that case. Assuming that
those calculations can be generalized to the present case in the most naive way, one can
argue that the emission rate \emmampdns\ depends on the energy $E$ of the ``closed string''
as follows \SahakyanCQ:
\eqn\nonpert{I_{LST}(E)\sim e^{-\pi\sqrt{k} E}~.}
The high energy density of states of LST is
\eqn\highelst{\rho_{LST}(E)\sim e^{2\pi\sqrt{k} E}~.}
This density of states is non-perturbative; it is obtained from considerations based on
black hole thermodynamics. Assuming that  the LST states that couple to the $D$-brane are again
``left-right symmetric'' we conclude that we need to consider $I_{LST}(E)\sqrt{\rho_{LST}(E)}$ summed
over all energies \SahakyanCQ.  As in the previous subsection, the exponentials in \nonpert, \highelst\
cancel. To calculate the typical energy of the LST states emitted by the brane one needs to
evaluate the corrections to \nonpert, \highelst, which are not known at present. One expects
this distribution of energies to be peaked around the energy of the $D0$-brane, as before.

\newsec{Tachyon DBI description of a BPS $D$-brane}

In the previous section we saw that the TDBI action captures some aspects of
the dynamics of non-BPS $D$-branes. In this section we will see that it can be
used to describe BPS $D$-branes as well. The basic observation \SenNF\ is that
since the potential \genv\ has two vacua, at $T\to\pm\infty$, one can attempt
to construct a kink solution $T=T(y)$, which interpolates between the two vacua
as a function of a spatial coordinate, $y$.

Analyzing the equations of motion of \simptach\ one finds that the kink has
vanishing width, \ie\ the solution has the form
\eqn\formkink{T(y)=\cases{-\infty&$y<y_0$\cr
                          +\infty&$y>y_0$\cr
                          }~.}
Clearly, this solution is singular, going from $-\infty$ to $+\infty$ instantaneously in $y$.
In order to study its properties, Sen \SenTM\ proposed to regularize \formkink\ to a field
configuration that interpolates between the two vacua smoothly in $y$ but does not solve the
equations of motion. As a parameter controlling the width of the kink is sent to zero, the
kink becomes more localized, and the violation of the equations of motion decreases. In the
limit, one finds the singular field configuration \formkink. Of course,
physical features of the resulting singular kink must not depend on the regularization.

The above description was used in \SenTM\ to demonstrate that the low energy effective action
on the kink is precisely the DBI action on a BPS $D$-brane. We will not review this calculation
here, since we will soon obtain the result in a simpler way. Here we note that the fact that
the kink solution \formkink\ is singular suggests that we should not think of it as a semiclassical
lump, like one does for non-singular soliton solutions in other contexts.

In the $D/NS$ system the two types of branes are identical, only differing in their orientation in
space \KutasovCT: the BPS $D$-brane wraps the ``tachyon'' circle, while the non-BPS one is
localized on it. It is natural to expect that this might be the case in flat $9+1$ dimensional
spacetime as well. Indeed, we will next show that this is a property of the TDBI action \effact,
which describes both systems.

To do that, note that the action \effact\ with the potential \genv\ has a solution in which the tachyon
field $T$ depends on a spatial coordinate, $y$, as follows:
\eqn\soleuc{\sinh\alpha T=A\cos\alpha y~.}
This is a Euclidean version of the rolling tachyon solution \rolling, but its
physical interpretation is rather different. For $A=0$ it describes a non-BPS
$D$-brane stretched in $y$. For generic $A$ the brane traces a more general curve
in the $(y,T)$ plane. As $A\to\infty$ it approaches a sequence of vertical lines
stretched in the $T$ direction, connected at large $T$ by brane segments that carry
no energy, due to the vanishing of the potential as $T\to\pm\infty$. These vertical lines
are localized in $y$, and correspond to (alternating) BPS $D$-branes and antibranes.

The solutions labeled by $A$, all of which have the same energy,\foot{To make the
energy finite, it is convenient to compactify the $y$ direction on a circle of radius
$n/\alpha$, for some integer $n$.} have a natural interpretation in the
full open string theory --
they correspond to the line of fixed points obtained by adding to the worldsheet
Lagrangian corresponding to a non-BPS $D$-brane the marginal boundary
perturbation\foot{Recall that $\alpha=1/\sqrt2$ for this case.}
\eqn\margopen{\delta\CL_{\rm ws}=\lambda\cos\left({y\over\sqrt2}\right)~.}
Varying $\lambda$ corresponds to interpolating between Neumann and Dirichlet
boundary conditions for $y$, \ie\ between a non-BPS $D$-brane stretched in $y$, and
an array of BPS branes and antibranes separated by $\delta y=\pi/\alpha=\sqrt{2}\pi$,
\CallanUB. The former corresponds in terms of \soleuc\ to $A=0$; the latter to $A\to\infty$.

Thus, we see that the TDBI action provides a natural description of BPS $D$-branes.
The solution \soleuc\ suggests that the tachyon direction should be treated on the
same footing as spatial directions such as $y$. The tachyon potential \genv\ can be
thought of \eg\ as due to a $T$-dependent dilaton. A BPS $D$-brane is one that is stretched
in $T$, while a non-BPS brane is localized in this direction, at the maximum of the
potential, $T=0$. Thus, the two appear on the same footing, as expected from the $D/NS$
system \KutasovDJ.

Another important point is that starting from the array of $D$ and $\bar D$-branes, which
corresponds to $A\to\infty$ in \soleuc, one can think of the line of solutions labeled
by $A$ as corresponding to condensation of the $D-\bar D$ tachyon, which is exactly
massless for a brane and antibrane at the critical separation $\sqrt{2}\pi$. Thus, we
see that the tachyon DBI action provides a description of the $D-\bar D$ system
that includes the tachyon, and gives a geometric picture of its condensation. We will
return to this issue in section 5.

The fact that BPS and non-BPS $D$-branes appear on the same footing as solutions of
the TDBI action \effact\ can be summarized as follows. Consider the action
\eqn\tdbiA{S=-\int d^{p+1}\xi e^{-\Phi}V(T)
\sqrt{-\det (G_{ab}+2\pi F_{ab})}~,}
where $\xi^a$, $a=0,1,\cdots, p$, are the worldvolume coordinates on the $D$-brane,
the open string metric is given by
\eqn\tdbiB{G_{ab}=g_{mn}\d_a x^m\d_b x^n+
\d_a T\d_b T~,}
and $g_{mn}(x)$ and $\Phi(x)$ are the spacetime metric and dilaton
(we will mostly take them to be trivial below). We have again suppressed terms
involving the fermions on the brane.

The action \tdbiA\ is invariant under reparametrizations $\xi^a\to f^a(\xi)$. It describes a 
$p$-brane propagating in a $10+1$ dimensional spacetime with metric $(m,n=0,1,\cdots, 9)$
\eqn\spmet{ds^2=g_{mn}(x)dx^mdx^n+dT^2~.}
The non-BPS $D$-brane is obtained by choosing the ``static gauge''
$x^a=\xi^a$, $a=0,1,\cdots, p$. The fields that enter the action \tdbiA\
in this gauge are the worldvolume gauge field, transverse scalars, and
tachyon. Their action is given by \effact, \defaa.

The BPS $D$-brane is obtained by choosing the static gauge $x^a=\xi^a$, $a=0,1,\cdots, p-1$,
$T=\xi^p$. Now, the tachyon direction is a coordinate along the brane; thus the $p$-brane
gives rise to a $(p-1)$-brane in the physical, $9+1$ dimensional, spacetime. The fields that
enter the gauge fixed action are the transverse scalars and
worldvolume gauge field. All these fields depend on all worldvolume
coordinates, including $T$, and the gauge field has a polarization in the $T$ direction, $A_T$.
Ignoring global issues, one can set $A_T=0$ by a $T$-dependent gauge transformation,
and we will assume below that this has been done. This leaves a residual gauge symmetry of
$T$-independent gauge transformations, which gives rise to spacetime gauge invariance on the
brane.

The above discussion is reminiscent of what happens in higher dimensional field theory
compactified on a circle. Indeed, as discussed in \KutasovCT, for some purposes
one can think of the tachyon direction as being compact. Thus, to analyze the low energy
dynamics of the BPS $D$-brane one can take the fields to be independent of $T$.
The bosonic part of the action takes now the form
\eqn\ssdbi{S=-\int d^pxdT V(T)\sqrt{-\det A}~,}
where ($\mu,\nu=0,1,\cdots, p-1$)
\eqn\defaa{A_{\mu\nu}=\eta_{\mu\nu}+
\partial_\mu y^I\partial_\nu y^I+2\pi F_{\mu\nu}~.}
Since the fields are independent of $T$, one can integrate over it. Using
the potential \vvv\ and defining
\eqn\taubps{
\tau^{BPS}_{p-1}\equiv\int_{-\infty}^\infty dTV(T)=\sqrt{2}\pi\tau_p~,
}
one arrives at
\eqn\ssddbbi{S=-\tau^{BPS}_{p-1}\int d^px\sqrt{-\det A}~,}
the standard DBI action on a BPS $D$-brane, with the correct \SenNF\
relation between the tensions of BPS and non-BPS $D$-branes \taubps.

It is easy to include the worldvolume fermions on the $D$-brane in the above
discussion. In type IIA string theory we define \SenNF,
\eqn\IIASDBI{\eqalign{
&\Pi^m_a =\d_a x^m-{\bar\theta}\Gamma^m\d_a \theta,\qquad
{\cal G}_{ab}=\eta_{mn}\Pi^m_a\Pi^n_b,
\cr
&{\cal F}_{ab}=2\pi F_{ab}-\left[{\bar\theta}\Gamma_{11}\Gamma_m\d_a \theta(\d_b x^m-
\frac{1}{2}{\bar\theta}\Gamma^m\d_b \theta)-(a\leftrightarrow b)
\right],\cr
&{\cal A}_{ab}=\d_a T\d_b T+{\cal F}_{ab}+
{\cal G}_{ab}~.
}}
Here $\theta$ is a single non--chiral Majorana spinor in ten
dimensions and $\Gamma_m$ are ten--dimensional Dirac matrices.
Then, the extension of the action \tdbiA\ that includes the fermions is
\SenTM
\eqn\SUSYDBI{
S=-\int d^{p+1}\xi V(T)\sqrt{-{\rm det}{\cal A}}+\int V(T)dT\wedge~
\left(\sum_{q\in 2Z+1}C^{(q)}\right)\wedge e^{\cal F}.
}
In a flat background, which we are considering here, the differential forms
$C^{(q)}$ are
\eqn\MixedFormC{
C^{(q)}=\frac{1}{q!}dx^{m_1}\wedge \dots \wedge dx^{m_{q-1}}\wedge
{\bar \theta}P_q\Gamma_{m_1\dots m_{q-1}}d\theta~,
}
where $P_q$ is equal to $\Gamma_{11}$ for $q=4k+1$ and to the unit matrix
for $q=4k+3$.

As explained above, a non-BPS $Dp$-brane extended in the directions
$(x^0,\cdots x^p)$ is obtained by fixing the static gauge $x^a=\xi^a$,
$a=0,1,2,\cdots, p$. Recall that the dimension of the non-BPS $D$-brane,
$p$, must be odd in the IIA case \SenNF.

To describe a BPS $D(p-1)$-brane extended in $(x^0, x^1,\cdots, x^{p-1})$,
we choose the static gauge $x^a=\xi^a$, $a=0,1,2,\cdots, p-1$, $T=\xi^p$.
Taking all the worldvolume fields to be independent of $T$ and integrating
over it, as before, we arrive at the standard action for a $D(p-1)$--brane
in IIA string theory \agan,
\eqn\SUSYDBIA{
S=-\tau^{BPS}_{p-1}\int d^{p}x \sqrt{-{\rm det}{\cal A}}+\tau^{BPS}_{p-1}\int
\left(\sum_{q\in 2Z+1}C^{(q)}\right)\wedge e^{\cal F},\quad
{\cal A}_{\mu\nu}={\cal F}_{\mu\nu}+
{\cal G}_{\mu\nu}~.
}
The tension $\tau^{BPS}_{p-1}$ is given by \taubps, as before.

The action \SUSYDBI\ contains a  thirty two component spinor $\theta$, but for a
BPS $D$-brane some components of this spinor turn out to be non-dynamical. To
see this, we expand \SUSYDBI\ to quadratic order in the fermions $\theta$ and
look at their kinetic terms.  For vanishing electromagnetic field, we find
\eqn\QuadrFerm{\eqalign{
S_{\rm ferm}=&\int d^{p+1}\xi V(T)\sqrt{-{\rm det}{G}}G^{ab}\d_a x^m
{\bar\theta}\Gamma_m\d_b \theta\cr
+&\frac{1}{p!}
\int V(T)dT\wedge dx^{m_1}\wedge\dots \wedge dx^{m_{p-1}}
\wedge
{\bar \theta}P_p\Gamma_{m_1\dots m_{p-1}}d\theta,\cr
}
}
where $G_{ab}$ is given by \tdbiB. To describe a BPS brane, which is stretched
in the $T$ direction, we partially fix the gauge $T=\xi^p$. Focusing on the massless
modes, we further take the coordinates $x^m$ and fermions $\theta_\alpha$ to be 
independent of $T$. Integrating \QuadrFerm\ over this variable, we arrive at the
quadratic action
\eqn\QdrKapProj{
S_{\rm ferm}=\tau^{BPS}_{p-1}\int d^{p}\xi \sqrt{-{\rm det}{G}}
G^{ab}{\bar\theta}(1+P_p\Gamma)\gamma_a\d_b \theta~,
}
where
\eqn\wwww{
G_{ab}=\eta_{mn}\d_a x^m\d_b x^n,\quad
\gamma_a=\d_a x^m \Gamma_m,\quad
\Gamma=\frac{1}{\sqrt{-G}}\gamma_{01\dots (p-1)}~.
}
Notice that, depending on $p$, the matrix $\Gamma$ squares to $1$ or to $-1$.
To arrive at equation \QdrKapProj\ we used the identity
$$
\frac{1}{p!}
d\xi^{a_1}\dots d\xi^{a_p}\gamma_{a_1\dots a_{p-1}}\d_{a_p}\theta=d^{p}\xi \sqrt{-{\rm det}{G}}G^{ab}\Gamma
\gamma_a\d_b\theta~.
$$
Since $P_p\Gamma$ in \QdrKapProj\ squares to one and is traceless, half of its eigenvalues are $+1$ 
and half are $-1$. Thus, only sixteen of the thirty two components of $\theta$ have a non-vanishing 
kinetic term and are dynamical. Although we only looked at the quadratic approximation \QdrKapProj, this
property persists to all orders in the fields $\theta$; it is a consequence of the local $\kappa$-symmetry
of the action \refs{\agan,\Refkappaone}.

The truncation of fermionic degrees of freedom from thirty two to sixteen, that happens for
$D$-branes which are uniformly stretched in the tachyon direction, does not occur for non-BPS $D$-branes,
which are localized in $T$. Indeed, in that case the second line in \QuadrFerm\ gives a cubic coupling
of two fermions and the tachyon, rather than a contribution to the kinetic term of $\theta$. The quadratic
Lagrangian for the fermions comes from the first line of \QuadrFerm\ and is non-trivial for all thirty two
components of $\theta_\alpha$.

The above discussion can be repeated for type IIB string theory. The dimension of
the non-BPS $D$-brane, $p$, is now even, and \IIASDBI\ is replaced by
\eqn\IIBSDBI{\eqalign{
&\Pi^m_a =
\d_a x^m-{\bar\theta}{\hat\Gamma}^m\d_a \theta,\qquad
{\cal G}_{ab}=\eta_{mn}\Pi^m_a\Pi^n_b,
\cr
&{\cal F}_{ab}=2\pi F_{ab}-\left[{\bar\theta}\tau_3
{\hat\Gamma}_m\d_a \theta(\d_b x^m-\frac{1}{2}{\bar\theta}
{\hat\Gamma}^m\d_b \theta)-(a\leftrightarrow b)
\right],\cr
&{\cal A}_{ab}=\d_a T\d_b T+{\cal F}_{ab}+
{\cal G}_{ab}~.
}}
Here $\theta$ consists of a pair of ten--dimensional Majorana--Weyl spinors:
\eqn\IIBPair{
\theta=\left(\eqalign{&\theta_1\cr &\theta_2}\right),
}
and the matrices entering \IIBSDBI\ are defined by
\eqn\IIBGamma{
{\hat\Gamma}^m=\left(\eqalign{\Gamma^m&\ \ 0\cr 0\ \ &\Gamma^m}\right),\qquad
\tau_3=\left(\eqalign{{\bf 1}&\ \ \ 0\cr 0\  &-{\bf 1}}\right).
}
The action \SUSYDBI\ is replaced by
\eqn\SUSYDBIIB{\eqalign{
S=&-\int d^{p+1}\xi V(T)\sqrt{-{\rm det}{\cal A}}+\int V(T)dT\wedge~
\left(\sum_{q-\rm{even}}C^{(q)}\right)\wedge e^{\cal F},\cr
&C^{(q)}=
\frac{1}{q!}dx^{m_1}\wedge \dots \wedge dx^{m_{q-1}}\wedge
{\bar \theta}P_q\Gamma_{m_1\dots m_{q-1}}d\theta~,
}}
where $P_{4k}=i\sigma_2$, $P_{4k+2}=\sigma_1$.
The rest of the discussion is the same as for the IIA case.

Another simple generalization is to the $D/NS$ system, where the potential
$V(T)$ is given by \vdnscircle. Starting with the reparametrization
invariant action \tdbiA, fixing the static gauge corresponding to a BPS
$D$-brane (\ie\ one which is wrapped around the circle transverse to the
fivebranes), and taking the worldvolume fields on the $D$-brane to be
independent of $T$, leads to the action \ssddbbi, with the tension of
the BPS brane given by the integral \taubps,
\eqn\taubpsdns{\tau^{BPS}_{p-1}=\sqrt{k}\pi\tau_p~,}
in agreement with the exact result \KutasovCT.

The reduction of the number of fermionic degrees of freedom due to
$\kappa$-symmetry for a brane stretched in the $T$ direction occurs
here as well. The $D$-branes used in the construction of \KutasovCT\
are BPS in $9+1$ dimensions. Thus, a $D$-brane localized on the circle 
transverse to the fivebranes has a sixteen component fermionic field living 
on its worldvolume, while for a $D$-brane wrapping the circle the analog of 
the $\kappa$-symmetry discussed above eliminates half of the fermions, 
and leaves a dynamical eight component spinor on the brane.

This agrees with spacetime expectations. It is well known from
the Hanany-Witten construction \HananyIE\ that a $D$-brane ending on a
stack of $NS5$-branes has the property that the massless fields on it
are the worldvolume gauge field and fields related to it by supersymmetry.
Since the $D/NS$ system preserves eight supercharges, those fields form
a vector multiplet of $N=1$ supersymmetry in $5+1$ dimensions, reduced
to the $p$-dimensional worldvolume of the $D$-brane. This multiplet contains
an eight-component fermion, in agreement with our discussion above.

\newsec{$T$-dependent dynamics on a BPS $D$-brane}

In the previous section we saw that
from the point of view of the TDBI action \tdbiA\
a BPS $D$-brane is one which is stretched in $T$.
We also saw that taking all the fields living on the $D$-brane to
be independent of $T$ leads to the usual, supersymmetric, DBI action
on the brane. The purpose of this section is to analyze the dynamics
of field configurations that depend on $T$.

The action \tdbiA\ describes a number of fields on a BPS $D$-brane.
We will focus for simplicity on a single scalar field $y(x^\mu, T)$
that describes the position of the $D$-brane in one of the transverse
directions. It is easy to generalize the discussion to other fields,
and we will comment on that later.

Choosing the static gauge corresponding to the BPS $D$-brane in \tdbiA,
and focusing on the dynamics of $y(x^\mu,T)$, leads to the action
\eqn\actyy{S=-\int d^pxdTV(T)\sqrt{1+(\partial_\mu y)^2+(\partial_T y)^2}~,}
where $V(T)$ is given as before by \genv. We are interested in solutions
$y(x^\mu, T)$ which satisfy the boundary conditions,
\eqn\boundc{\lim_{T\to\pm\infty} y(x^\mu, T)=0~,}
corresponding to localized (in $T$) perturbations of the BPS $D$-brane.

We divide our analysis of the dynamics of $y$ into
two parts. In subsection 4.1 we discuss small fluctuations $y(x^\mu, T)$
described by the quadratic action
\eqn\quadrAct{
S=-\frac{1}{2}\int d^p x dTV(T)\left[(\partial_\mu y)^2+(\partial_T y)^2\right]
}
obtained by expanding \actyy\ to leading order in $y$. We will see that despite
the fact that the effective length of the $T$ direction $\int dTV(T)$ is finite
(see \taubps), the equation of motion of \quadrAct\ does not have normalizable
Kaluza-Klein type modes. Generic initial waveforms move to large $|T|$ and grow with time.
Eventually, they enter the regime where the non-linearities of the action \actyy\ become
important. In subsection 4.2 we study this regime by a combination of analytic and
numerical tools. In subsection 4.3 we interpret the results. Subsection 4.4 contains some
remarks on other modes of the TDBI action.

\subsec{Linear analysis}

The Euler-Lagrange (EL) equation for the  action \quadrAct\ is
\eqn\eullag{\partial_T\left[V(T)\partial_Ty\right]+V(T)\partial_\mu\partial^\mu y=0~.}
To look for normalizable modes, we separate variables,
\eqn\sepvar{y(x^\mu, T)=\phi(x^\mu) z(T)~.}
For a mode with mass $m$, one has
\eqn\phimm{\eqalign{
&\partial_\mu\partial^\mu\phi =m^2 \phi~,\cr
&\d_T\left[V(T)\d_Tz\right]+m^2V(T)z=0~.
}}
The equation for $z(T)$ on the second line of \phimm\
is analyzed in appendix A.1, where it is shown that
it has no normalizable solutions. Potential divergences
come from $T\to\pm\infty$; the only case for which
the contributions of both regions are finite is $m=0$, $z(T)=$
const, which corresponds to the zero mode discussed in the
previous section.

Since \phimm\ does not give rise to normalizable modes, we go
back to the EL equation \eullag\ and ask how generic perturbations
of the BPS $D$-brane that are localized in $T$ (and thus satisfy
the boundary conditions \boundc) evolve in time. For simplicity
we consider the case with no dependence on the spatial coordinates
along the $D$-brane, $x^i$, so $y=y(t, T)$. This analysis is applicable
to a $D0$-brane, or the zero momentum mode on a higher dimensional brane.

For any initial waveform $y(0,T)$, $\dot y(0,T)$, one can in principle
solve the EL equation \eullag\ and find $y(t,T)$ for later times. A generic
waveform splits into two parts, which move in the positive and negative $T$
directions respectively, and reach in a finite time the region where $|T|$ is
sufficiently large that we can replace $\cosh\alpha T$ in the potential
\genv\ by $\exp(\alpha|T|)$. We will restrict the discussion to this region,
which as we saw before, is the interesting one for the normalizability
analysis. Without loss of generality we will take $T$ to be positive.

In this region, the EL equation \eullag\ takes the form:
\eqn\LinearIn{
y''-\ddot y-\alpha y'=0~.
}
Here prime denotes a derivative w.r.t. $T$, while dot stands for a time derivative.
As is familiar from studies of scalar fields in linear dilaton backgrounds (which
is what the action \quadrAct\ describes for large $T$ \KutasovDJ), it is
convenient to rescale $y$ as follows:
\eqn\rescyy{y=e^{\alpha T/2}w~.}
In terms of $w$, \LinearIn\ takes the form
\eqn\MassKG{w''-\ddot w-m^2 w=0,\quad m\equiv \frac{\alpha}{2}~,
}
a massive Klein-Gordon equation with mass $\alpha/2$. The general solution of this equation
is described in appendix A.2. Here we present a qualitative picture that is sufficient for our
purposes.

The solution of \MassKG\ is a combination of left and right-moving waves.
We are interested in the right-moving one, which can be written as\foot{One needs to take
the real part of this expression, as usual.}
\eqn\rightmove{w(t,T)=\int_0^\infty dp\omega(p)e^{i\left[pT-E(p)t\right]}~, }
where $\omega(p)$ is the Fourier transform of the initial waveform $w(t=0,T)$, and
\eqn\eepp{E(p)=\sqrt{p^2+m^2}~.}
The different momentum components of \rightmove\ move to the right (\ie\ in the positive
$T$ direction) with different velocities, $v(p)=p/E(p)$. The high momentum modes, $p\gg m$,
propagate with a speed close to that of light. Thus, if we write the initial
waveform $w(0,T)$ as
\eqn\decww{w(0,T)=w_{\rm low}(T)+w_{\rm high}(T)~,}
where $w_{\rm high}$ corresponds to momenta much higher than $m$ and $w_{\rm low}$
corresponds to the rest, at later times the high momentum part of the waveform has the 
(approximate) form $w_{\rm high}(T-t)$, while the low momentum part lags behind and is 
more complicated.

\ifig\linFig{Evolution of $w(t,T)$, \rescyy, which satisfies the
Klein-Gordon equation \MassKG, and $y(t,T)$, which satisfies the
equation \LinearIn, presented at time $t=0$,\ $t=10$, and $t=1000$.}
{\epsfysize5in\epsfbox{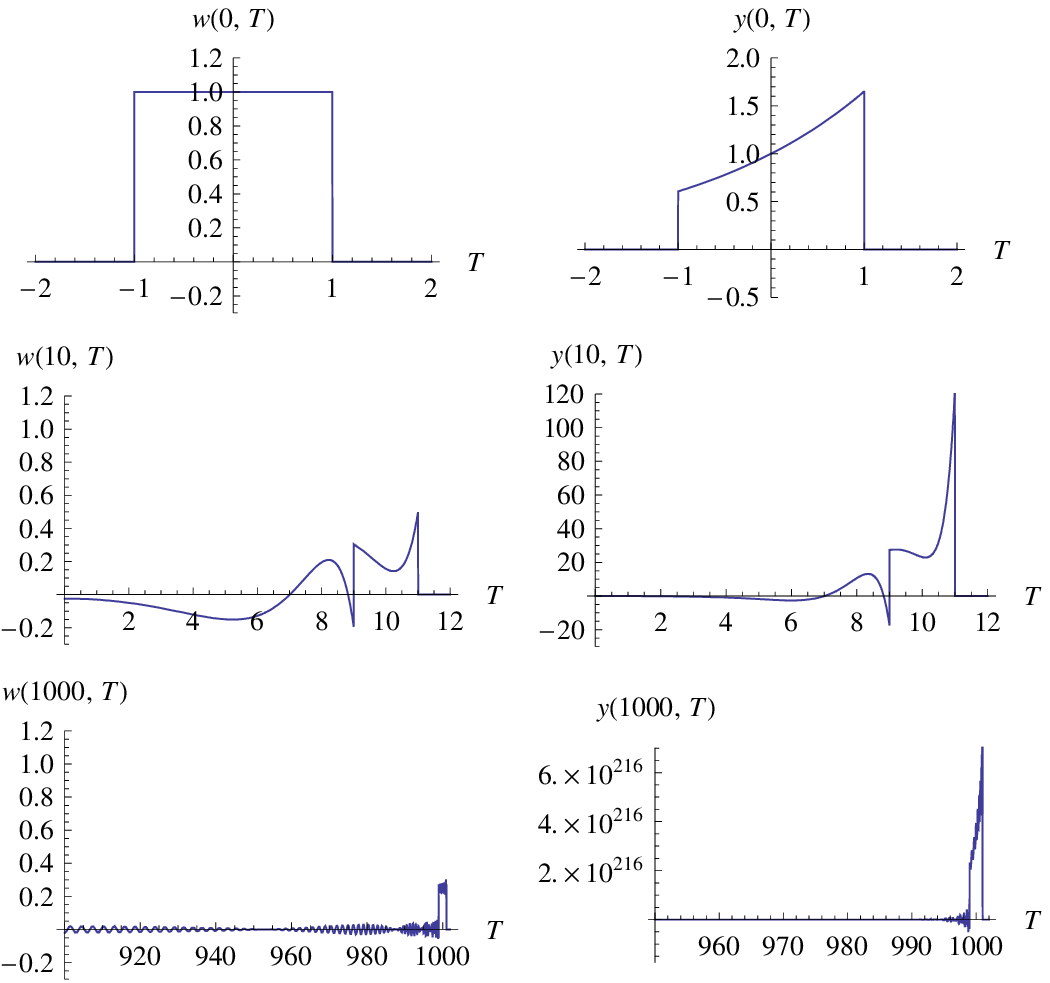}}

As time goes by, the different momentum components separate in space
due to dispersion, with the highest momenta at the leading edge of
the wave. As a result, the waveform oscillates more and more rapidly
in space. In \linFig\ we present numerical results on the time
evolution of a particular initial waveform, which illustrates the
above discussion.

We see that $w$ propagates to large $T$ while roughly preserving its
height; hence, $y$, which is related to it by the rescaling \rescyy\
grows exponentially with $T$. Since the leading edge of the waveform
propagates with the speed of light in the $T$ direction, the largest
$y$ along the waveform grows roughly like
\eqn\maxyy{y_{\rm max}\sim e^{\alpha t/2}~.}
Eventually the linear approximation breaks down, and one needs to take into account the
non-linearities of the action \actyy.

\subsec{Nonlinear analysis}

When the waveform $y(t,T)$ studied in the previous subsection increases beyond the regime
of validity of the linear approximation \LinearIn, we have to go back to the full TDBI action \actyy,
\eqn\ExpAct{
S=-2\tau_p\int dt dT e^{-\alpha T}\sqrt{1-\dot y^2+(y')^2}~,
}
where we assumed that $\alpha T\gg 1$ as before. The EL equation of \ExpAct,
\eqn\FullNL{
\d_T\left[\frac{e^{-\alpha T}y'}{\sqrt{1-{\dot y}^2+(y')^2}}\right]=
\d_t\left[\frac{e^{-\alpha T}\dot y}{\sqrt{1-{\dot y}^2+(y')^2}}\right]~,
}
is non-linear. Deviations from linearity are characterized by the function
\eqn\defS{
F=1-{\dot y}^2+(y')^2.
}
When $F(t,T)$ is close to one, \FullNL\ reduces to \LinearIn, but in general it is
more complicated. For localized perturbations which satisfy the boundary conditions
\boundc, $F$ is close to one far from the center of the waveform, while close to the
center it deviates from one by an amount that grows with time. This is where the
linear approximation breaks down for late times.

A related problem was studied in \refs{\FelderSV,\FelderXU}, who considered
the action \simptach\ for the tachyon on a non-BPS $D$-brane extended in a spatial
direction $y$. For large $T$ this action takes the form
\eqn\KofAct{
S=-2\tau_p\int dt dy e^{-\alpha T}\sqrt{1+(\d_y T)^2-(\d_t T)^2}~.
}
The actions \ExpAct\ and \KofAct\ are related by a change of variables\foot{This can
be thought of as a special case of the discussion of the reparametrization invariant action
\tdbiA\ in section 3.} from $T$ to $y$. The form \ExpAct\ is more useful for expanding around a brane
stretched in $T$ \boundc, while \KofAct\ is better for expanding around a brane stretched in $y$,
such as those studied in \refs{\FelderSV,\FelderXU}.

The authors of \refs{\FelderSV,\FelderXU} found that generic initial configurations
$T(y, t=0)=T_0+\delta T(y)$, where $T_0$ is constant and $\delta T(y)$ is a small smooth
$y$-dependent perturbation, lead at late time to configurations for which the argument of the
square root in \KofAct,
\eqn\Pkof{
P\equiv 1+(\d_y T)^2-(\d_t T)^2~,
}
vanishes everywhere. For $\delta T(y)=0$ this is indeed a property of the (exact) rolling tachyon
solution \roldns. For other initial conditions one finds more general ($y$-dependent) solutions of \Pkof.

The discussion of \refs{\FelderSV,\FelderXU} cannot be applied to our case directly, due to the different
boundary conditions \boundc. Far from the center of the growing waveform $y(t,T)$ the
linear approximation \LinearIn\ is good and the Jacobian of the change of variables from $y$ to $T$,
$|y'|$, is singular. However, close to the center of the waveform this Jacobian is finite and the two
problems are locally similar. In particular, the two functions \defS\ and \Pkof\ are simply related,
\eqn\fprel{F=(y')^2 P~.}
Thus, it is natural to expect that in regions where $y'$ is finite, $F(t,T)$
approaches zero at late times.

To see that this is plausible, consider the following toy problem. Start at $t=0$ with
a $D$-brane which forms a straight line in the $(T,y)$ plane,\foot{Of course, such a $D$-brane does not
satisfy the boundary conditions \boundc.}
\eqn\PreStrBrn{
y(0,T)=aT~,}
and is at rest, $\dot y(0,T)=0$. The equation of motion \FullNL\ can be solved exactly for this case:
\eqn\StraigtBrane{
y(t,T)=aT-\frac{1+a^2}{a\alpha}\ln\cosh\frac{a\alpha t}{\sqrt{1+a^2}}~.
}
This solution has a simple interpretation. The exponential potential in \ExpAct\ provides a force
in the positive $T$ direction. Since the brane \PreStrBrn\ is tilted, only the component of the
force transverse to the brane acts on it. This causes the brane to accelerate in the transverse
direction. Its velocity in this direction is given by
\eqn\instvel{v(t)=\tanh{a\alpha t\over\sqrt{1+a^2}}~.}
At late times, $t\gg \frac{\sqrt{1+a^2}}{a\alpha}$, \instvel\ approaches the speed of light,
\StraigtBrane\ behaves like
\eqn\StraigtBraneAs{
y=aT-\sqrt{1+a^2}~t~,
}
and the function $F$ \defS\ approaches zero.

Since $y'=a$ is finite, one can change coordinates from $T$ to $y$, and consider
\StraigtBrane\ as a solution of the tachyon DBI action \KofAct. In fact, this configuration
interpolates between a BPS $D$-brane, which corresponds to $a=0$, and the rolling tachyon
solution ($a\to\infty$).

Going back to the evolution of waves $y(t,T)$ on a BPS $D$-brane in
the non-linear regime, the above discussion suggests the following
qualitative picture. The exponential potential in \ExpAct\ provides
a force on the perturbed brane towards large $T$. Under the influence
of this force the waveform $y(t,T)$ accelerates and at late time
approaches the speed of light in the $(T,y)$ plane. The quantity $F$,
defined in \defS, approaches zero in part of the core of the waveform,
where the analysis of \refs{\FelderSV,\FelderXU} applies.

\ifig\NLFig{Evolution of $y(t,T)$, which satisfies equation \FullNL,
and the corresponding $F(t,T)$ defined by \defS, presented at time
$t=0$,\ $t=4$, and $t=5.5$. After time $t=5.5$ the numerical
integration breaks down.} {\epsfxsize5in\epsfbox{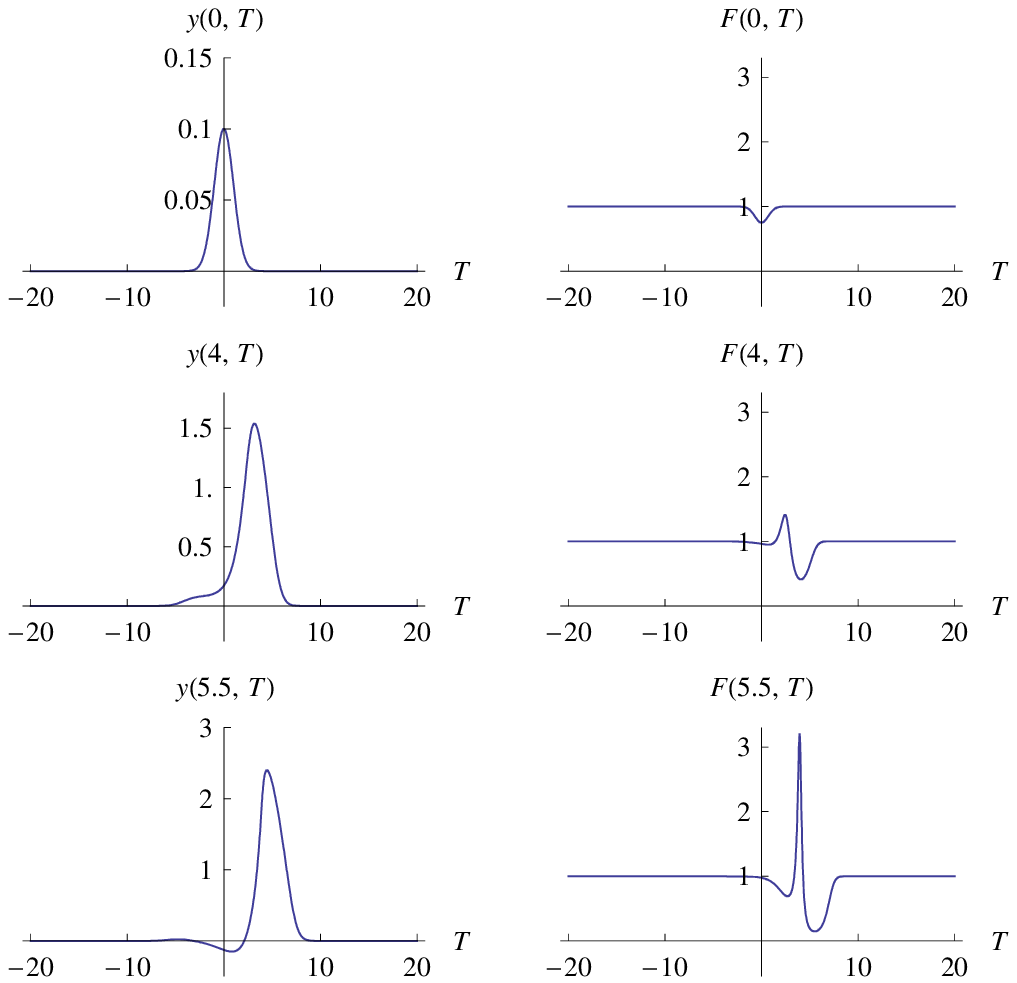}}

To verify the above picture, we numerically solved the EL equation
\FullNL\ for a few typical initial conditions. We present the results in 
\NLFig. The numerical evolution breaks down when $F$ \defS\ becomes 
too small, but the qualitative features of the solution up  to that point can 
be seen to agree with  expectations. It would of course be nice to 
understand the behavior of the solutions of \FullNL\ better.

\subsec{Boundary interpretation}

In this and the previous sections we saw that the TDBI action leads to a description of BPS $D$-branes
in critical string theory in $\IR^{9,1}$ as objects extended in an extra dimension labeled by $T$,
and studied the dynamics associated with this dimension. In this subsection we interpret
the results in terms of the usual $9+1$ dimensional spacetime of critical string theory.

We saw that small perturbations which are localized in $T$ and initially localized in $y$ very close
to the $D$-brane, extend to larger $y$ while moving to large $T$ as time goes by. If we suppress the
$T$ direction, the picture we have in $9+1$ dimensions is the following. We start with a BPS $D$-brane
localized in $y$, add an energy of order $1/g_s$, initially localized near the brane\foot{From the TDBI 
point of view, this energy is contained in the perturbation $y(t,T)$, which we assume to be small but finite 
as $g_s\to 0$.} and evolve it in time. The brane emits two ``jets'' in the $\pm y$ directions, which carry the 
excess energy, while remaining localized in all other transverse directions. Each jet carries non-zero 
momentum $p_y$, also of order $1/g_s$; the total momentum depends on the initial perturbation.\foot{The
above picture is modified by $g_s$ effects, which lead to radiation of light
closed string modes, expanding the jets in the transverse directions at late times.}

At first sight the above picture seems puzzling: the mode we excited seems to be an open string
mode (since it is described by the TDBI action), but all such modes are expected to be localized on the
brane, whereas the mode we studied propagates to large $y$ at late times. In fact, generalizing the logic
of section 2 to this case, it is not difficult to understand what is going on.

We saw in section 4.1 that $T$-dependent perturbations $y(t,T)$ of the TDBI action do not give
normalizable states. Thus, they do not describe open string excitations. Rather, starting with
a particular initial waveform $y(0,T)$ corresponds to creating a non-trivial closed string state
with energy of order $1/g_s$. As discussed in the introduction, such states are expected to be
described by classical open string theory; in this case, the open string description is in terms
of the TDBI action.

As time goes by, the closed strings move away from the $D$-brane in the $y$ direction. Their motion,
stress-energy distribution, etc, can be calculated from the TDBI action. The fact that
these closed strings form sharply defined jets which are moving out in $y$ but are localized in
the other directions is directly related to the fact that their energy is of order $1/g_s$.

All this is in accord with the open string completeness proposal mentioned in section 2. The
system analyzed here actually provides a sharper example of this proposal than the original context
in which it was made (the rolling tachyon system). There, the closed strings described by the
TDBI action remained localized near the location of the original non-BPS brane, while here they
move away from it. Thus, it is clear that the TDBI action describes a closed string excitation
of the BPS $D$-brane.

Our discussion also provides an illustration of the idea that one can think of the tachyon direction
in space as an analog of the holographic direction in gauge-gravity duality, \ie\ as a direction that
encodes the energy of physical processes. Indeed, displacing a localized waveform $y(t,T)$ by
$T\to T+\delta T$, changes its energy in the way familiar from holography, with $T\to\infty$ corresponding
to the infrared.

So far we discussed the interpretation of our TDBI analysis for non-BPS excitations of BPS $D$-branes
in critical string theory. As we saw earlier in the paper, this analysis is also applicable to the
$D/NS$ system, which provides further support to the picture presented in this subsection.

A BPS $D$-brane is obtained in this case by wrapping a $D$-brane around the circle
transverse to the fivebranes. The tachyon direction parameterizes position on this circle. From the
point of view of an observer living on the fivebranes, such a $D$-brane is a defect which is localized
in some of the directions in $\IR^5$. For example, in type IIB string theory we can consider a D-string
wrapping the circle, which looks like a pointlike defect in $\IR^5$.

Perturbing the shape of the D-string by turning on a non-vanishing  $y(0,T)$ for one of the
transverse scalars can be thought of as exciting the defect by a background of ``closed strings,'' or
modes in the boundary theory, that couple to it. The TDBI action describes the dynamics of these modes
as they move away from the defect.

This can be qualitatively understood as follows. A $D$-string ending on an $NS5$-brane corresponds
in $5+1$ dimensional terms to a particle charged under the gauge field on the worldvolume of the
fivebrane. It can be described, following \CalMal, as a spike solution of the fivebrane gauge theory.
Introducing a localized perturbation $y(t,T)$ corresponds in this language to a deformation of the
gauge field and transverse scalars on the fivebrane worldvolume. These fields are the analogs of closed 
strings in $5+1$ dimensional LST, and their deformations are analogous to non-trivial closed string 
backgrounds. Thus we see that in the $D/NS$ system, TDBI perturbations localized in $T$ indeed 
correspond to ``closed string'' modes in the boundary theory.

The above description is only qualitative since IIB LST reduces to a gauge theory only at low energies,
while the region where the TDBI action is applicable corresponds to high energies. In that regime one has
to treat the $D$-string as an excitation of the full LST. 

\subsec{Other modes}

In addition to translational modes, such as $y$ \actyy, the TDBI action of a BPS $D$-brane \SUSYDBI\
describes other fields, such as the worldvolume gauge field $A_\mu$. The $T$-independent mode of
$A_\mu(x^\nu,T)$ gives rise to the massless gauge field on the BPS brane, as explained in section 3.
For the $T$-dependent modes one can proceed as for the scalar $y$ above.

For small $A_\mu$ one can expand the action and solve the Maxwell equations for a perturbation localized
in $T$. As usual, one can choose a gauge in which the equation of motion for $A_\mu$ is the same as for
the scalar modes, \LinearIn. Thus, electric and magnetic fields on the brane tend to grow with time, and
their location in $T$ increases as well. For late times one needs to take into account the non-linearities
of the TDBI action. The analog of the fact that the function $F$ \defS\ approaches zero at late times for
some values of $T$ around the core of the scalar waveform $y(t,T)$ is that the electric field approaches
its critical value there.

As for the scalar mode, one can think of the $T$-dependent gauge field as describing a very
massive closed string excitation created in the vicinity of the brane. Unlike the situation
there, as the electric and magnetic fields grow the excitation remains localized near the brane
in the transverse directions. This is similar to what happens for the rolling tachyon background
discussed in section 2.

\newsec{TDBI description of the $D-{\bar D}$ system}

In this section we study the brane-antibrane system. In string theory, a parallel $Dp-\bar Dp$ pair
separated by a distance $L$ along a spatial direction $y$ has the property that the lowest lying mode
of a string stretched between the brane and the antibrane (the $D-\bar D$ tachyon) is tachyonic when
\PolchinskiRR
\eqn\llccrr{L<L_{\rm cr}=\pi\sqrt2~.}
It is massive for $L>L_{\rm cr}$ and massless for $L=L_{\rm cr}$. We will see that this
is the case for the TDBI description of these branes as well. We will also discuss the
coupling of the tachyon to other modes on the branes.

\subsec{$D-\bar D$ tachyon}

To describe the $D-\bar D$ tachyon using the TDBI action, we recall the discussion of the
brane configuration \soleuc. We saw in section 3 that as the parameter $A$ goes to infinity,
this configuration approaches a sequence of equidistant $D$ and $\bar D$-branes at the
critical distance, $L_{\rm cr}$. These branes are stretched in the $T$ direction, with opposite
orientations for branes and antibranes. Taking $A$ to be finite corresponds in string
theory to turning on a non-zero expectation value of the (massless) $D-\bar D$ tachyon.

\ifig\DDbartach{The BPS $D$ and $\bar D$-branes depicted in (a) can be thought of as connected
at $T=\infty$. This brane configuration can be continuously deformed into that of (b), (c).
For $L>L_{\rm cr}$, (b), there is a force pushing it back to  (a), while for $L<L_{\rm cr}$, (c),
it is pushed away.}
{\epsfysize2.5in\epsfbox{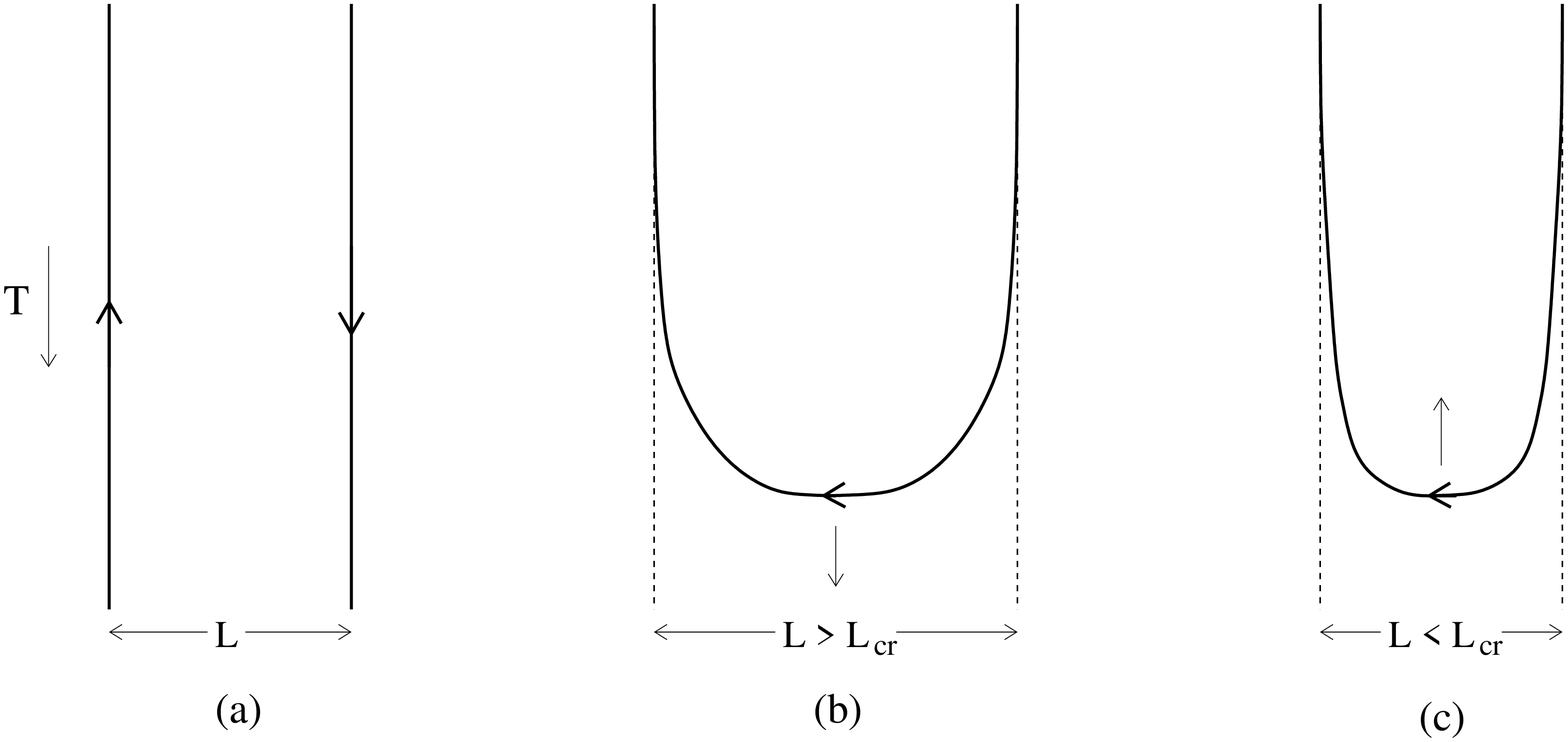}}

Thus, we conclude that $D-\bar D$ tachyon condensation corresponds
in the TDBI description to a process in which the brane and
antibrane connect at large $T$. This process may seem discontinuous,
since it requires a brane and an antibrane which are a finite
distance apart to connect. In fact, one can view the $D-\bar D$
configuration as containing a brane segment which connects the
branes at $T=\infty$. The extra segment does not carry energy due to
the fact that the tension of the brane goes to zero at large $T$
(see \tdbiA). In the process of tachyon condensation the brane is
deformed as in \DDbartach. It is a dynamical question whether
such a deformation increases the energy, in which case the
corresponding mode is massive, or decreases it (in which case it is
tachyonic). We next show that the answer depends on $L$, as in
string theory.

Since in the TDBI description the $D-\bar D$ tachyon is localized at large $T$,
to study the onset of its condensation we can focus on this region, \ie\ replace
the full TDBI action by \ExpAct\ (or, equivalently, \KofAct). We start with a $D$
and $\bar D$-brane, which are stretched in $T$ and localized in $y$ a distance
$L$ apart. To regularize the divergence in the energy from $T\to-\infty$ we
impose the constraint $y(T=T_0)=\pm L/2$, with the large and negative constant
$T_0$ playing the role of a cutoff.

The regularized energy of the disconnected $D-\bar D$ pair (\DDbartach(a)) is given by
\eqn\edisc{E_{\rm disc}=2\int_{T_0}^\infty dT e^{-\alpha T}={2\over\alpha}e^{-\alpha T_0}~.}
We would like to show that for $L>L_{\rm cr}=\pi/\alpha$, any small deformation of the
branes of the sort depicted in \DDbartach(b) increases the energy, while for $L<L_{\rm cr}$
the energy can be continuously decreased by such deformations (\DDbartach(c)).

Consider first the case
\eqn\largel{L>L_{\rm cr}={\pi\over\alpha}~,}
and examine a one parameter sequence of smooth static branes, which satisfy
the boundary conditions. One can, for example, take the parameter to
be the maximal value of $T$ reached by the U-shaped $D$-brane. Suppose
one can decrease the energy of the $D-\bar D$ pair by the
reconnection process of \DDbartach(b). Then, we can
choose the sequence of configurations mentioned above such that the
energy of the U-shaped brane continuously decreases when the branes
reconnect. As the maximal value of $T$ decreases, eventually this
energy must increase again. E.g., for a brane that is stretched in
the $y$ direction at fixed $T=T_0$, one has \KofAct
\eqn\straightb{E_{\rm conn}=L e^{-\alpha T_0}~,}
which is larger than \edisc\ in the regime under discussion \largel. Thus, there must be a configuration
between the two extremes \edisc, \straightb, that minimizes the energy, and therefore is a static solution
of the equations of motion of \ExpAct, \KofAct. However, the only smooth static solution of these equations
is the $T\to\infty$ limit of \soleuc,
\eqn\hairpp{e^{\alpha T}=2A\cos\alpha y~,}
the hairpin solution of \LukyanovNJ, which is irrelevant in the
regime \largel. Therefore, we conclude that it must be that the
initial assumption, that the energy can be continuously decreased by
the reconnection process of \DDbartach(b) is
incorrect, and in fact, the $D-\bar D$ configuration of
\DDbartach(a) is a local minimum of the energy functional.

For $L<L_{\rm cr}$, the same discussion leads to the opposite conclusion. Now, there is
a static solution of the equations of motion with the right boundary conditions,
\hairpp\ with
\eqn\formaa{A={e^{\alpha T_0}\over2\cos{\alpha L\over2}}~.}
Since this solution is unique, it is clear from the arguments above that it must be
a local minimum of the energy functional. This can be verified directly by computing
its energy
\eqn\enmin{E_{\rm conn}={1\over\alpha A}\tan{\alpha L\over2}=
{2\over\alpha}e^{-\alpha T_0}\sin{\alpha L\over2}~,}
where in the last equality we used  \formaa. Comparing \enmin\ to \edisc,
we see that the former is smaller, as expected.

Thus, we conclude that the brane deformation of \DDbartach\ gives rise to massive
modes for $L>L_{\rm cr}$, while for $L<L_{\rm cr}$ the lowest lying mode corresponding
to such deformations is tachyonic. This is in agreement with what one finds in string theory.

\ifig\DDbarenergetics{For $L=L_{\rm cr}$, the deformation that takes the $D-\bar D$ pair (a)
to a connected brane with the shape \hairpp, (b), is massless. The energy of the brane
is independent of the parameter $A$ in \hairpp.} {\epsfysize2.5in\epsfbox{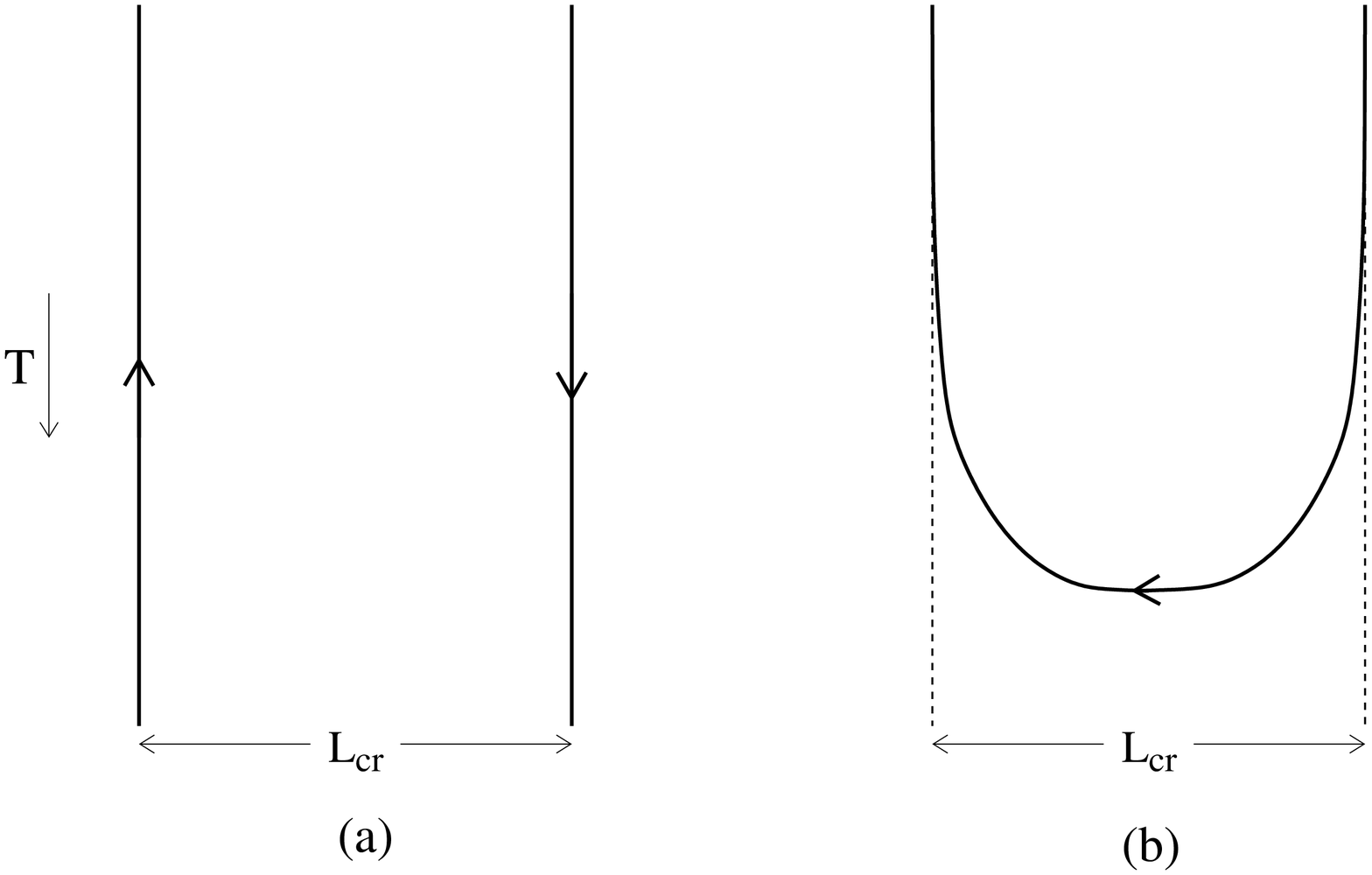}}

For $L=L_{\rm cr}$ the ``tachyon'' is massless. Its expectation value
is related to the constant $A$ in \hairpp, and the shape of the brane is
depicted in \DDbarenergetics. To construct the low energy
effective action for it, one promotes this constant to a spacetime
field,\foot{It is natural to take $A(x^\mu)$ to be independent of
$y$  since it represents a massless mode, but this assumption can be
justified by a more careful analysis.} $A=A(x^\mu)$. Substituting the
ansatz $\exp(\alpha T)=2A(x^\mu)\cos\alpha y$ into the TDBI action \ExpAct,
gives rise to the following effective action for $A(x^\mu)$:
\eqn\ssaaa{S=-4\tau_p\int d^p x dT{A e^{-\alpha T}\over\sqrt{4A^2-e^{2\alpha T}}}
\sqrt{1+\left(\partial_\mu A\over2\alpha A^2\right)^2e^{2\alpha T}}~.}
For slowly varying $A$  one can expand the square root in \ssaaa. The
first two terms in the derivative expansion are
\eqn\sstwo{S_2=-4\tau_p\int d^p x dT \left[
{A e^{-\alpha T}\over\sqrt{4A^2-e^{2\alpha T}}}+{1\over8\alpha^2}
{e^{\alpha T}(\partial_\mu A)^2\over A^3\sqrt{4A^2-e^{2\alpha T}}}\right]~.}
The second term in \sstwo\ is a kinetic term for $A$. The integral over $T$
in it is convergent, and can be evaluated, yielding
\eqn\ssttkin{S_k=-{\pi\tau_p\over2\alpha^3}\int d^p x{(\partial_\mu A)^2\over A^3}~,}
where we took into account the factor of two in the integral over $T$ coming from
the two arms of the hairpin.

It is convenient to define
\eqn\canphi{\Phi ={2\over\alpha} A^{-\half}~,}
in terms of which one has (using \taubpsdns)
\eqn\kincan{S_k =  - {\tau_{p-1}^{BPS}\over2}\int d^p x \partial_\mu \Phi
\partial^\mu \Phi~.}

\noindent
The first term in \sstwo\ is a potential for $A$. It is divergent from $T\to-\infty$,
due to the infinite energy of the straight part of the hairpin in \DDbarenergetics(b),
where the connected brane is essentially identical to the $D-\bar D$ pair of 
\DDbarenergetics(a). This divergence is  due to the fact that we restricted 
to the large $T$ regime, and is absent in the full problem, in which the potential 
$\exp(-\alpha T)$ is replaced by $1/(2\cosh\alpha T)$.

Since, in any case, we are only interested in the energy difference between the configurations
of \DDbarenergetics(a) and (b), which is finite, we can subtract from \sstwo\ the energy of
the configuration of \DDbarenergetics(a). To avoid manipulating infinite quantities we can
introduce the cutoff $T_0$ (as in the discussion around \edisc), subtract the resulting finite
expressions, and send $T_0\to-\infty$. This gives rise to the potential
\eqn\vvvaaa{V(A)=-4\tau_p\left[
\int_{-\infty}^{T_1} dT e^{-\alpha T}\left({2A \over\sqrt{4A^2-e^{2\alpha T}}}-1\right)
-\int_{T_1}^\infty dT e^{-\alpha T}\right]~,}
where $T_1$ is determined by $\exp(\alpha T_1)=2A$. The two integrals in \vvvaaa\ are
convergent; they can be calculated exactly and shown to cancel. Thus, the potential
$V(A)$ vanishes, in agreement with the fact that $A$ is a modulus for $L=L_{\rm cr}$.

We see that the two derivative action for $\Phi$ \canphi\ is just that for a free massless
field, \kincan. Expanding \ssaaa\ to higher orders gives rise to higher
derivative terms of the form
\eqn\highder{S=- {\tau_{p-1}^{BPS}\over2}\int d^p x \partial_\mu \Phi
\partial^\mu \Phi\sum_{n=0}^\infty a_n\left(\partial_\mu\Phi\partial^\mu\Phi\over\Phi^2\right)^n~,}
with  calculable constants $a_n$; \eg, $a_0=1$. Thus, the quadratic action \kincan\
provides a good description when expanding around a vacuum with non-zero $\langle\Phi\rangle$,
as in \DDbarenergetics(b). However, as $\langle\Phi\rangle\to 0$, \ie\ as we approach the
configuration of \DDbarenergetics(a), the range of validity of \kincan\ becomes smaller and
smaller, and at the origin this action fails.

It is natural to ask how the scalar field $\Phi$ is related to the $D-\bar D$ tachyon
in critical string theory, which is also massless for $L=L_{\rm cr}$. One obvious
difference between the two is that while the $D-\bar D$ tachyon is complex, the field
$\Phi$ \canphi\ is real. This is due to the fact that we restricted the TDBI discussion to 
large values of the coordinate $T$. We will see later that when we go back to the full problem,
with the potential \genv, we recover the full complex tachyon field.

Another apparent difference between the $D-\bar D$ tachyon and $\Phi$ is the structure
of their effective actions. We saw that the effective Lagrangian of $\Phi$, \highder, contains 
high order interactions, suppressed by the scale $\langle\Phi\rangle$. For the tachyon,
one expects similar interactions to be suppressed by the scale $l_s$; well below the
string scale, we expect the tachyon field to be free.

If the scale $\langle\Phi\rangle$ is large in string units, the corrections \highder\
to the free Lagrangian for $\Phi$ are small at low energies and can be neglected.
However, when that scale is well below the string scale, we seem to have a disagreement
between the two. Thus, we would like to propose that the TDBI description is useful for 
large $\langle\Phi\rangle$. When this vev decreases below the string scale, $\alpha'$
corrections to the TDBI action become important and the correct description goes over
to the one familiar from open string theory.  We will return to this relation below.

For $L<L_{\rm cr}$ the lowest lying $D-\bar D$ mode is tachyonic. This can
be seen directly by studying time-dependent solutions of the equations of motion of
\KofAct. An exact solution of these equation is
\eqn\timedepsol{T=at+{1-a^2\over\alpha}\ln\cos{\alpha y\over\sqrt{(1-a^2)}}~,}
which looks like a hairpin, with the two arms a distance
\eqn\llaa{L={\pi\over\alpha}\sqrt{1-a^2}=L_{\rm cr}\sqrt{1-a^2}} apart. Since this distance
is smaller than the critical one, \largel, the hairpin is unstable
to tachyon condensation, or in the present language to the
reconnection process of \DDbartach(c). The
configuration \timedepsol\ is a time dependent solution associated
with this instability.

For $a>0$ it describes a hairpin which approaches the $D-\bar D$
configuration of  \DDbartach(a) as $t\to\infty$. One
can think of this as a solution in which the $D-\bar D$ open string
tachyon, which is tachyonic in the regime \llaa, climbs up its
inverted quadratic potential, approaching the top as
$t\to\infty$. The solution with $a<0$ (and the same $L$ \llaa)
describes the time reverse process.

To recapitulate, we find that the TDBI description of the $D-\bar D$ system contains
a mode, which we refer to as the $D-\bar D$ tachyon, which is tachyonic
for $L<L_{\rm cr}$, massless for $L=L_{\rm cr}$ and massive for $L>L_{\rm cr}$.
The value of $L_{\rm cr}$, \largel, is equal to the one obtained in string theory.
In the large $T$ limit, where the potential \genv\ is exponential, the tachyon has
an exactly flat potential for $L=L_{\rm cr}$. Its condensation leads to the sequence
of solutions labeled by $A$ \hairpp.

In the above analysis, we have focused on the large positive $T$ region, and found that
it contains a scalar field with $L$-dependent mass. In this region, one of the two
real components of the tachyon, as well as the gauge fields and translational modes on
the $D$ and $\bar D$-branes are frozen. In order to incorporate them into the discussion
we need to return to the full problem, with potential \genv.

In this case, there are two asymptotic regions, $T\to\pm\infty$, each of which contains
a localized mode of the sort discussed above. These are the two components of the complex
$D-\bar D$ tachyon. The gauge fields and translational modes on the branes also become
normalizable. We will discuss them below.

In classical open string theory, the stability properties of the $D-\bar D$ tachyon are
directly related to those of the tachyon on a non-BPS $D$-brane (see \eg\ \refs{\SenNF,\HarveyNA}).
Consider a non-BPS $D$-brane wrapped around a circle of radius $R$ labeled by $y$.
The lowest non-zero momentum mode of the open string tachyon on the brane has $p_y=1/R$.
Condensing it corresponds on the worldsheet to adding the boundary interaction (compare
to \margopen)
\eqn\relopen{\delta\CL_{\rm ws}=\lambda\cos\left(y\over R\right)~.}
For $R>\sqrt2$ this perturbation is relevant or, equivalently, tachyonic in spacetime.
Thus, the non-BPS $D$-brane is unstable to tachyon condensation whose endpoint is a $D-\bar D$
pair at the distance $\pi R$. For $R<\sqrt 2$ the situation is reversed: the perturbation
\relopen\ is irrelevant on the worldsheet, \ie\ massive in spacetime. The non-BPS $D$-brane is
stable to this perturbation, while the $D-\bar D$ pair at the distance $\pi R$ is unstable to
the corresponding $D-\bar D$ tachyon condensation. For $R=\sqrt2$ the perturbation \relopen\
becomes exactly marginal. $\lambda$ parameterizes a line of fixed points smoothly connecting 
the non-BPS brane and the brane-antibrane pair, as discussed around eq. \margopen.

The TDBI analysis gives the same qualitative structure. The non-BPS $D$-brane is
stretched in $y$ and localized at the origin in $T$.
The tachyon field on it has the same mass as in string theory; thus, its lowest momentum
mode, which has $p_y=1/R$, is tachyonic, massless or massive in the same ranges of $R$ as
in the string theory analysis. In particular, for $R>\sqrt2$, it grows exponentially with
time, and approaches at late time a pair of branes stretched in $T$ a distance $\pi R$
apart -- the $D$ and $\bar D$-branes of the string theory analysis. Similarly, for
$R<\sqrt2$ the $D-\bar D$ configuration is unstable to the reconnection process described
above. It decays to the non-BPS $D$-brane configuration (if we fine-tune
the zero momentum tachyon on the non-BPS $D$-brane to zero).

In string theory, a $D-\bar D$ pair on a circle of radius $R$ actually has an infinite number
of modes of the $D-\bar D$ tachyon labeled by the winding around the circle. These
modes have natural TDBI counterparts. In the covering space, the brane configuration
looks like an infinite array of alternating branes and antibranes separated by the
distance $\pi R$. The $D-\bar D$ tachyon described above corresponds to a mode that connects
a $D$-brane to an adjacent $\bar D$-brane. However, we can also consider modes for which
a $D$-brane is connected to a $\bar D$-brane separated from it by $n$ $D-\bar D$
pairs. Such a mode carries winding number $n$ around the circle. It is easy to see that it
becomes massless at the same value of $R$ as in the full string theory analysis.

\ifig\Dwinding{Excitations of a $D$-brane winding around a circle give rise to TDBI
analogs of open strings that end on the brane and wind around the circle.}
{\epsfysize2.5in\epsfbox{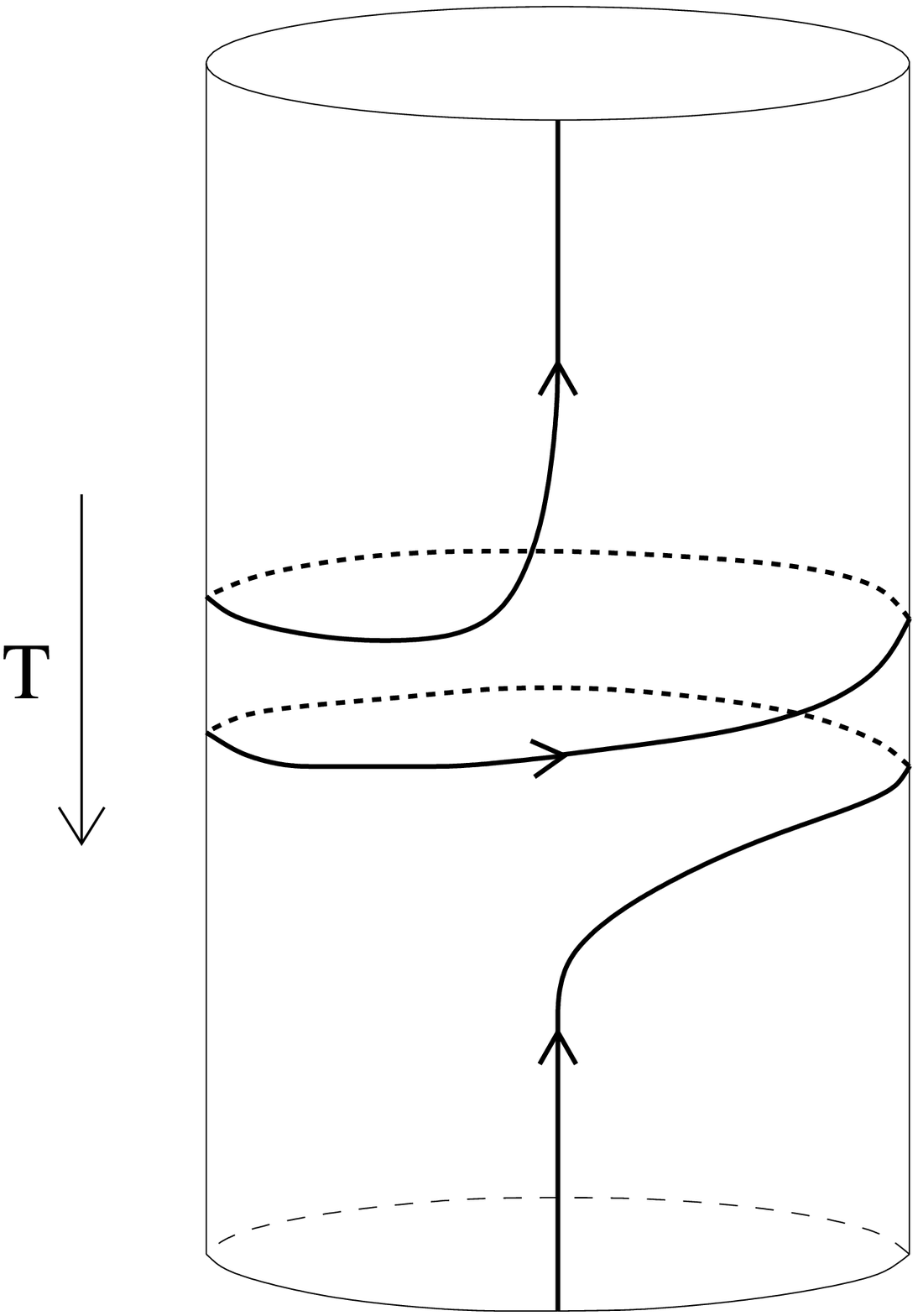}}

In string theory, similar winding strings give a tower of massive
modes corresponding to the light fields on a single $D$-brane, such as
the translational modes, gauge fields and fermions discussed in
section 3. From the TDBI point of view, these modes correspond to
perturbations of a $D$-brane that winds $n$ times around the circle as
$T$ varies between $-\infty$ and $\infty$ (see \Dwinding).

Another feature of the classical string theory analysis that is nicely realized in the TDBI
picture is the couplings of the tachyon to the massless fields on the brane and antibrane.
The couplings to the gauge fields on the branes will be discussed in the next subsection. Here
we comment on the coupling to the translational modes.

The mass of the tachyon depends on the separation between the brane and the antibrane
$\Delta y$ as follows:
\eqn\masstt{m^2=-\half+\left(\Delta y\over2\pi\right)^2~.}
This means that when the tachyon condenses, a potential is generated for the massless scalar
field that corresponds to the brane-antibrane separation. This potential, which is due to the
term proportional to $(\Delta y)^2 |T|^2$ in the energy, pushes $\Delta y$ towards zero whenever
$T\not=0$.

\ifig\DDbarcompact{$D-\bar D$ tachyon condensation (vertical arrows) leads to a potential for
the scalar field corresponding to the separation of the branes, pushing the separation towards
zero (horizontal arrows).}
{\epsfysize2.5in\epsfbox{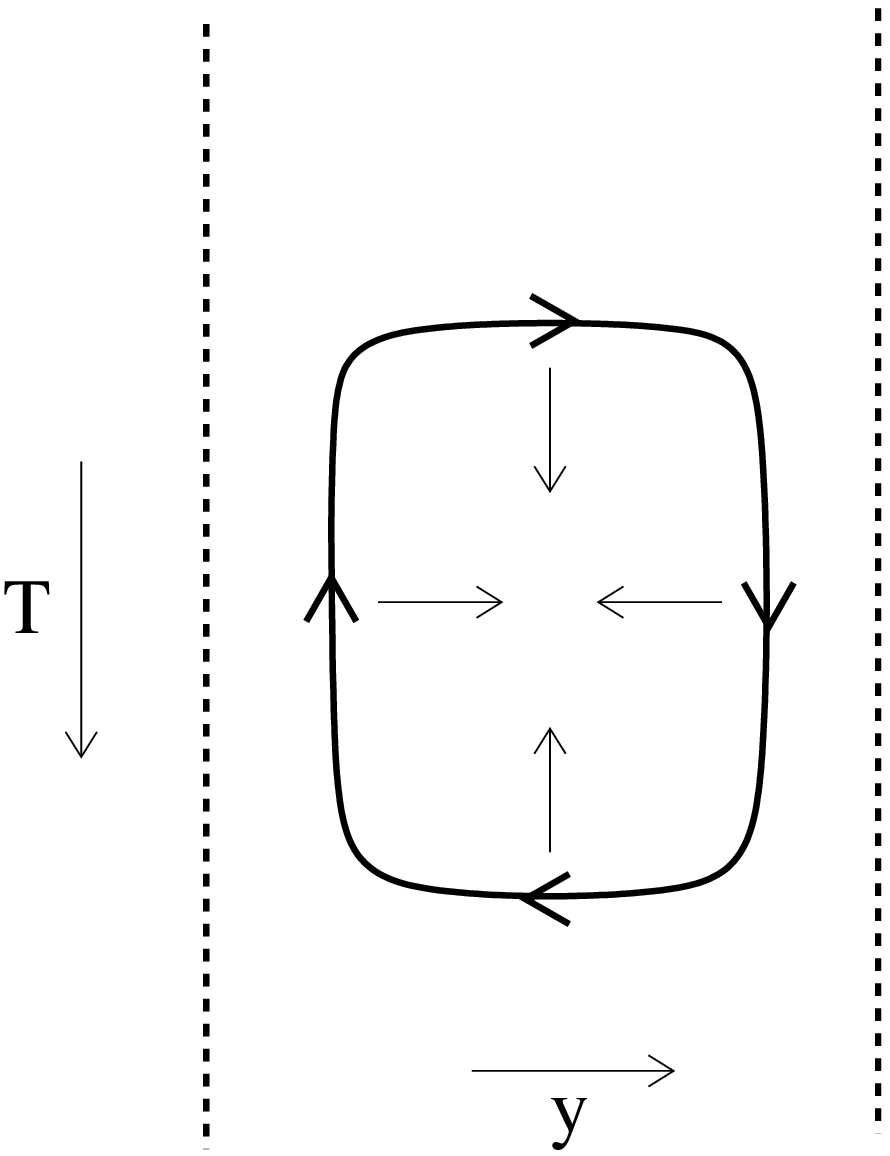}}

From the TDBI perspective, once the $D$ and $\bar D$-branes connect
and form a single connected brane (for $L<L_{\rm cr}$), it is energetically 
advantageous for them to shrink both in the $T$ and the $y$ directions 
(see \DDbarcompact). This is a reflection of the fact that both the tachyon 
and the relative translation modes
condense in this process. Complete annihilation of the branes corresponds
to the shrinking of the brane of  \DDbarcompact\ to a point. The last stage of this process
seems singular, but in fact is not. When the shrinking $D$-brane gets to
the regime where its energy is comparable to that of fundamental string
states, the $D$-brane description becomes unreliable, and is replaced by a
perturbative string one.

At first sight, the above discussion seems inconsistent with the existence of
the TDBI solution \soleuc. As explained above, this solution corresponds to a
$D-\bar D$ pair on a circle of circumference $2L_{\rm cr}$. The brane and antibrane
are separated by the critical distance $L_{\rm cr}$ and, for finite $A$, the
tachyon condensate is non-zero. One can ask why there is in this case no tadpole
for the mode corresponding to the brane -- antibrane separation.

From the string theory point of view, since the $D-\bar D$ pair is on a circle,
there are now two tachyon fields, $\CT_L$ and $\CT_R$, corresponding to fundamental
strings stretched between the $D$ and $\bar D$-branes,  to the left and to the
right of one of them. The potential for the relative position mode $\Delta y$ is
now\foot{Up to an overall multiplicative constant.}
\eqn\arraypot{V=\left[-\half+\left(\Delta y\over2\pi\right)^2\right]|\CT_L|^2
+\left[-\half+\left(\sqrt{2}-{\Delta y\over2\pi}\right)^2\right]|\CT_R|^2~.}
Expanding around the critical separation,
\eqn\critsep{\Delta y=\sqrt{2}\pi(1+\chi)~,}
we find
\eqn\stablepot{V=\half|\CT_L|^2\left(\chi^2+2\chi\right)+
\half|\CT_R|^2\left(\chi^2-2\chi\right)~.}
For generic $\CT_L$, $\CT_R$, \stablepot\ gives rise to a tadpole for the relative
separation $\chi$, but for
\eqn\equallr{\CT_L=\CT_R=\CT~,}
this tadpole vanishes, and the relative separation modulus becomes massive.
All this is in agreement with the TDBI picture, as discussed in the next subsection.

\subsec{Gauge fields}

In section 3 we saw that the TDBI action \tdbiA\ for a BPS $D$-brane, \ie\ one which is stretched
in the $T$ direction, gives rise to a massless gauge field, corresponding to the constant ($T$
independent) mode of the worldvolume gauge field $A_a$. Thus, for a $D-\bar D$ pair it gives
two gauge fields, on the brane and antibrane, respectively.

When the $D-\bar D$ tachyon discussed in the previous subsection is turned on, one expects the
difference between the $D$ and $\bar D$ gauge fields, under which the tachyon is charged, to
become massive, with a mass proportional to the vev of the tachyon. The diagonal gauge field
should remain massless.

In order to describe this phenomenon, we start with the action \simptach\ and potential \genv,
and consider the static solution of its equation of motion, \soleuc. As explained in section 3,
it describes a $D-\bar D$ pair at the critical distance \largel, corresponding to $A\to\infty$
in \soleuc, with a finite condensate of the $D-\bar D$ tachyon, which is massless in this case.
This condensate is labeled by $A$, which can be thought of as the zero mode of a massless 
scalar field $A(x^\mu)$.

In order to study the interaction between the tachyon and the gauge field, we need to include
the latter in the action \simptach. This is easily done by going back to the full action \tdbiA,
and fixing the static gauge. This leads to
\eqn\TDBIfull{S = -\tau_p\int dy d^p x \frac{1}{\cosh(\alpha T)}
\sqrt{(1+T'^2)(1+2\pi^2 F_{\mu \nu} F^{\mu \nu}) + 4\pi^2 F_{\mu
y}{F^\mu}_y+ \partial_\mu T \partial^\mu T}~.}
Plugging in the solution \soleuc\ with $A=A(x^\mu)$, and expanding to quadratic order in
$\partial_\mu A$, $F_{\mu\nu}$ leads to
\eqn\TDBITin{\eqalign{ S = - \tau_p\int dy d^p x [
&\pi^2\frac{\sqrt{1+A^2}}{1+A^2\cos^2(\alpha y)}F_{\mu \nu}F^{\mu
\nu} + 2\pi^2\frac{F_{\mu y}{F^\mu}_y}{\sqrt{1+A^2}} \cr
+&\frac{1}{2\alpha^2}\frac{\partial_\mu A \partial^\mu A \cos^2(\alpha
y)}{\sqrt{1+A^2}(1+A^2\cos^2(\alpha y))} ]~.\cr}}
In the tachyon kinetic term (the second line of \TDBITin), the integral over $y$ can be
performed, yielding
\eqn\TachKin{ S_T = -\tau_p\int d^p x
\frac{\pi}{\alpha^3}\frac{\partial_\mu A \partial^\mu
A}{A^2(1+A^2)}(\sqrt{1+A^2}-1)~.}
For large $A$ (\ie\ for small deviations from the $D-\bar D$ configuration),
\TachKin\ takes the form
\eqn\TachKinexp{S_k = - \tau_p\int d^p x
\frac{\pi}{\alpha^3}\frac{\partial_\mu A \partial^\mu A}{A^3}~.}
Note that this is larger by a factor of two from our previous result \ssttkin.
The reason for the apparent discrepancy is that in \TachKinexp\ we are describing
a perturbation that connects the brane and antibrane both at large positive $T$
(say to the left of the brane) and at large negative $T$ (to the right of it).
In the notation of \arraypot, we are turning on both $\CT_L$ and $\CT_R$ (see
\equallr). On the other hand, in the derivation of \ssttkin\ we focused on only
one of these perturbations, by restricting to large positive $T$. Thus, \ssttkin\
and \TachKinexp\ in fact agree.

Defining
\eqn\Avars{ \phi =2{\sqrt{2\pi\tau_p\over A\alpha^3}}}
brings \TachKinexp\ to canonical form
\eqn\TachKincan{S_k =  - \half\int d^p x \partial_\mu \phi\partial^\mu \phi~.}
The first line of \TDBITin\ gives rise to kinetic terms for the gauge fields.
As $A\to\infty$, we expect them to reduce to separate kinetic terms for the
gauge fields on the brane and antibrane. To see that this is
indeed the case, note that the second term (proportional to $F_{\mu y}^2$) goes
to zero in this limit. To evaluate the first term we use the relation
\eqn\limaadel{\lim_{A\to\infty}\frac{\sqrt{1+A^2}}{1+A^2\cos^2(\alpha y)}=L_{\rm cr}
\left[\delta (y-{L_{\rm cr}\over2})+\delta (y+{L_{\rm cr}\over2})\right]~.}
Plugging this into \TDBITin\ we find that the gauge kinetic term is given in this limit by
\eqn\gaugekin{S_{gk}=-{1\over4}(2\pi)^2\tau_{p-1}^{BPS}\int d^px
\left[\left(F_{\mu\nu}^{(L)}\right)^2+\left(F_{\mu\nu}^{(R)}\right)^2\right]~,
}
where $F_{\mu\nu}^{(L)}$ and $F_{\mu\nu}^{(R)}$ are the gauge fields on the $D$-brane
(localized at $y=-L_{\rm cr}/2$) and $\bar D$-brane (localized at $y=L_{\rm cr}/2$),
respectively, and we used the relations \taubps, \taubpsdns, \largel,
\eqn\ddtt{\tau_{p-1}^{BPS}=\tau_p L_{\rm cr}={\tau_p\pi\over\alpha}~.}
To analyze the action \TDBITin\ for finite $A$, we  choose the gauge $A_y=0$, so the
dynamical fields are $A_\mu(x^\nu,y)$. We also replace the field $A(x^\mu)$, which
parametrizes the tachyon (see \Avars) by its expectation value, for simplicity.

The resulting equations of motion for the gauge field are
\eqn\EOM{\frac{\sqrt{1+A^2}}{1+A^2\cos^2(\alpha y)} \partial_\mu F^{\mu\nu}
+ \frac{\partial^2_y A^\nu}{\sqrt{1+A^2}}=0~.}
Separating variables,
\eqn\sepans{A^\mu(x_\nu,y) =\tilde{A}^\mu(x_\nu)\psi(y)~,}
and assuming that the gauge field $\tilde A_\mu$ has mass $m$,
$\partial_\mu \tilde{F}^{\mu \nu} = m^2\tilde{A}^\nu$,
leads to an eigenvalue equation for $\psi(y)$:
\eqn\EOMsep{\psi'' = -m^2\frac{1+A^2}{1+A^2\cos^2(\alpha y)}\psi~.}
One solution is $\psi={\rm const}$, which has $m=0$.  It corresponds to the diagonal
gauge field on the two $D$-branes, as can be seen by looking at the action \TDBITin.
The second term in this action, which involves $(\partial_y A_\mu)^2$, vanishes in this
case. The first term is sharply peaked at the locations of the brane and antibrane, due
to \limaadel, and corresponds to the sum of the two gauge fields, $A_\mu^{(L)}+A_\mu^{(R)}$.

Note that the constant mode is massless for all values of $A$, in agreement with the fact
that the $D-\bar D$ tachyon $\phi$ \Avars\ is uncharged under the diagonal gauge field. 
For $A=\infty$ (\ie\ for vanishing tachyon condensate) the action of the massless mode, 
\TDBITin, is supported entirely at $y=\pm L_{\rm cr}/2$, the locations of the brane and 
antibrane, due to \limaadel. For finite $A$ (\ie\ finite tachyon condensate) the wavefunction
of the massless mode spreads out in $y$.

To analyze the eigenvalue equation \EOMsep\ for non-zero $m$, it is convenient to restrict
to large $A$ (small tachyon vev \Avars). Defining $\mu^2 =m^2\sqrt{1+A^2}\simeq m^2 A$,
and using  \limaadel, \EOMsep\ becomes:
\eqn\Schro{\psi'' =
-\mu^2L_{\rm cr}\left[\delta(y-{L_{\rm cr}\over2}) +
\delta(y+{L_{\rm cr}\over2})\right]\psi~.}
This differential equation, with periodic boundary conditions chosen such that \soleuc\ contains
a single $D-\bar D$ pair, describes a particle in a Dirac comb with period $2L_{\rm cr}$
and zero energy. This gives
\eqn\musol{\mu^2 = {4\over L_{\rm cr}^2}~.}
The corresponding wavefunction, $\psi_1 (y)$, is odd under $y\to -y$.

Plugging back into the action \TDBITin\ we conclude that the second term in the action
gives rise to a mass term for the relative $U(1)$, $A_\mu^{(L)}-A_\mu^{(R)}$, of the form
\eqn\ssmass{S_m=-{1\over 4} m^2 (2\pi)^2\tau_{p-1}^{BPS}\int d^p x\left( A_\mu^{(L)}-A_\mu^{(R)}\right)^2~,}
where
\eqn\massaa{m^2=\frac{4\alpha^2}{\pi^2 A}={\alpha^4\phi^2\over2\pi^2\tau_{p-1}^{BPS}}~.}
The fact that the mass term \ssmass\ is proportional to $\phi^2$ is consistent
with the fact that $\phi$ is a component of a complex scalar field that is charged
under the relative $U(1)$ gauge field. We can read off the charge of the scalar
field by comparing to the standard field theoretic analysis of the abelian Higgs
model (see \eg\ \PeskinEV). This leads to the conclusion that the charge
of $\phi$ under $A_\mu^{(L)}$ and $A_\mu^{(R)}$ is $\pm e_\phi$,
\eqn\chargephi{e_\phi=\alpha^2~.}
This can be compared to the string theoretic analysis of the tachyon, which gives
\PolchinskiRR\ $e_T=1$.

At first sight it appears that the TDBI result \chargephi\ is half of the string theory
one (recall that $\alpha=1/\sqrt2$ for the critical string). However, we would like to
argue that it is in fact consistent. The key point is that, as we discussed
after eq. \highder, the TDBI analysis is valid for large values of the tachyon field,
$\CT\gg 1$ in string units. Thus, it is possible that the relation between the string
theory tachyon and $\phi$ \Avars\ is non-trivial. Our result \chargephi\ suggests that
the string theory tachyon $\CT$ is related to (the complex field whose real component
is) $\phi$ as follows:
\eqn\reltachphi{\CT\propto\phi^2~.}
Later we will see that this identification is consistent with other facts as well.

For the $D/NS$ system, \chargephi\ gives the suggestive result $e_\phi=1/k$, \ie\
the charge of $\phi$ is $1/k$ that of a fundamental string. This may be  related
to the duality between $NS5$-branes and $Z_k$ orbifolds, but we will leave a more
detailed discussion to future work.

In the previous subsection we saw that in the presence of a non-zero tachyon 
condensate, the relative translational mode $\chi$, \critsep, also becomes massive. 
In appendix B, we calculate its mass in the TDBI regime, and find that 
$m_\chi^2\propto\phi$.  This should be compared to the small tachyon regime,
$\CT\ll1$, where $m_\chi^2\propto|\CT|^2$, \stablepot.   

\subsec{Real time $D-\bar D$ dynamics and open-closed string duality}

Combining the discussion of section 4 and this section leads to an
appealing picture of real time brane-antibrane dynamics described by
the TDBI action.

\ifig\DDbarpulse{The TDBI description of real time brane -- antibrane interaction.}
{\epsfxsize5in\epsfbox{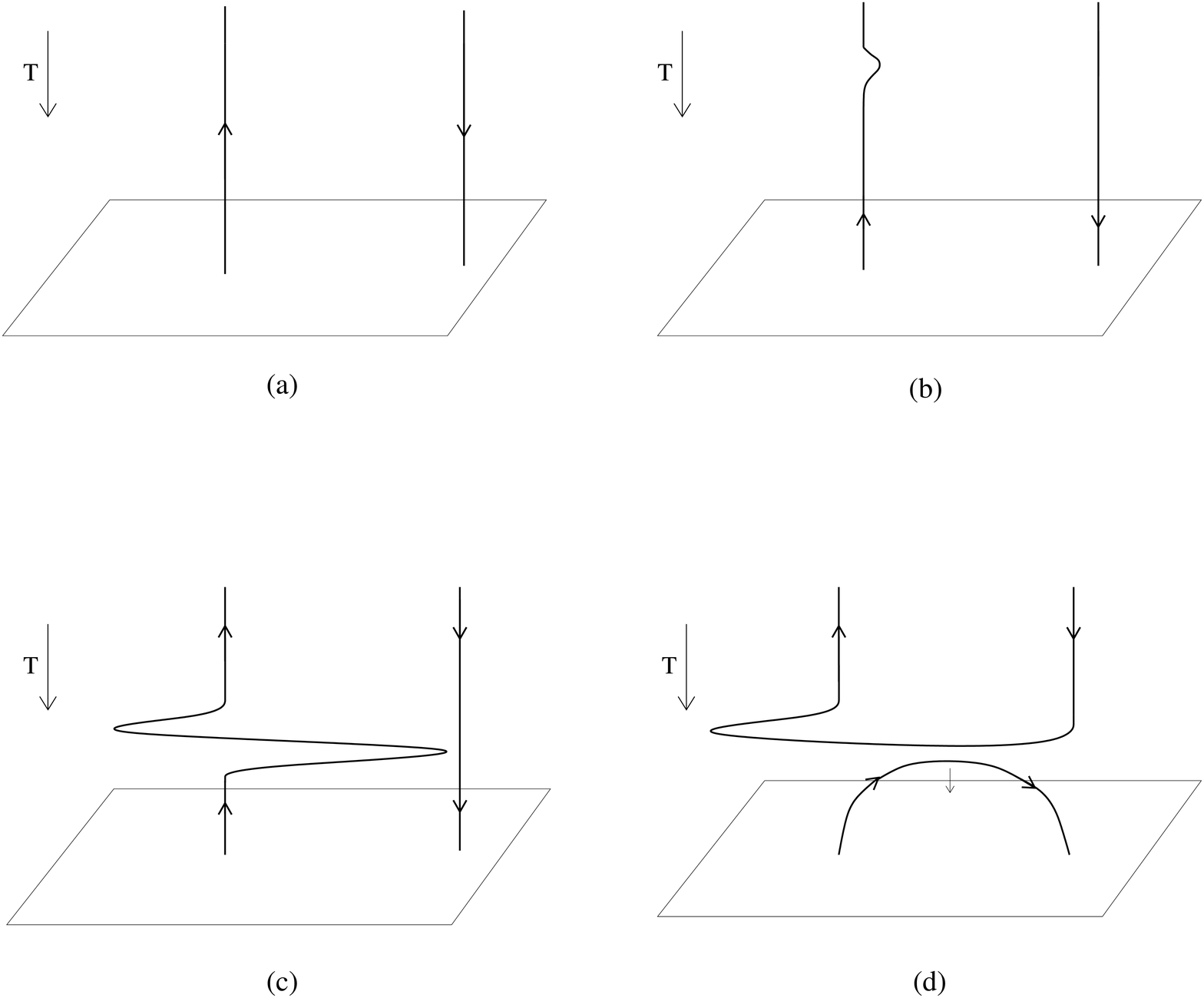}}

Consider a $D$-brane and a $\bar D$-brane separated by a distance
$L$ in the $y$ direction, as before. Both branes are stretched in the $T$ 
direction, with opposite orientations, as indicated in \DDbarpulse(a). 
Imagine that  a small perturbation of the scalar field  $y(x^\mu,T)$ localized 
in $T$ is introduced on one of the branes (see \DDbarpulse(b)). 
As we saw in section 4, such a perturbation splits into two parts, which 
propagate to large positive and negative $T$ respectively, and grow in 
$y$ in the process (see \DDbarpulse(c)).

At some point, the perturbation propagating towards $T\to\infty$
(say) reaches the other brane. The branes touch, and there is a
finite probability that they reconnect, as indicated in
\DDbarpulse(d). After the reconnection, the part of the branes
that is localized entirely at large $T$ falls towards $T\to\infty$
under the influence of the potential \genv, and disappears.

The remaining part looks like that discussed earlier in this section
in the context of $D-\bar D$ tachyon condensation. If the distance
between the $D$ and $\bar D$ branes is large, so that the $D-\bar D$
tachyon is massive, it collapses back to the straight $D-\bar D$
configuration, (as in the discussion of \DDbartach(b)). On the other
hand, for $L<L_{\rm cr}$ it eventually (after the brane smoothes out
and approaches the shape of \DDbartach(c), corresponding to the
tachyonic mode) follows the type of trajectory discussed before in
the context of tachyon condensation.

From the point of view of the discussion of section 4, the dynamical process described
above corresponds to an exchange of closed strings with energy of order $1/g_s$ by the
brane and antibrane. When the distance between the two $D$-branes is very large, this
exchange provides a small correction to the attraction between the branes due to exchange
of light closed strings (\ie\ gravitational attraction). However, as the brane separation
decreases the effect grows, and for $L<L_{\rm cr}$ it leads to brane-antibrane annihilation.

From the point of view of the discussion of this section, the above
dynamical process is associated with $D-\bar D$ tachyon
condensation. In particular, the reconnected brane of
\DDbarpulse(d) was identified above with a configuration in
which the $D-\bar D$ tachyon has a non-zero expectation value.

The two descriptions of the dynamics are related to each other via (the TDBI version of)
open-closed string duality. The usual open-closed string duality relates tree level closed
string exchange between $D$-branes to a one loop amplitude involving open strings stretched
between the branes. The reason the relation is between closed string tree level and open
string one loop is that the open and closed strings in question have energies that remain
finite in the limit $g_s\to 0$ (\ie\ they are perturbative string states).

On the other hand, the closed strings that play a role in our discussion have energies of
order $1/g_s$. The exchange of such strings in the closed string channel is equivalent to
a classical open string effect -- the condensation of the $D-\bar D$ tachyon. This is another
manifestation of  Sen's open string completeness proposal discussed above. It makes it
clear that this proposal can be thought of as a generalization of open-closed string duality
to energies or order $1/g_s$.

\newsec{D1--D3 system}

In the previous sections we have seen that the extra dimension of space labeled by $T$
plays an important role in the dynamics of non-BPS $D$-branes and $D-\bar D$ systems. 
Its properties are very reminiscent of those of the holographic direction in bulk -- 
boundary duality. 

We discussed two types of systems. The first is non-supersymmetric brane systems
in $9+1$ dimensional string theory. There, the ``boundary'' theory is the full 
ten dimensional critical string theory, while the ``bulk'' theory is not well 
understood beyond the TDBI approximation. The second involves 
$D$-branes in the background of $NS5$-branes. In this case, the bulk theory
is ten dimensional string theory in the near-horizon geometry of the fivebranes,
while the boundary theory is the $5+1$ dimensional Little String Theory.\foot{The 
compactness of one of the directions transverse to the fivebranes can be neglected 
at  large $T$.} This theory reduces to standard field theory at low energies, but at high 
energies it is non-local and is not well understood.    

To gain further insight into different elements of our discussion, it might be
useful to study the analogous construction in a case where similar issues arise,
but both the bulk and the boundary theories are under control. In this subsection 
we discuss such a system. 

Consider a stack of $N$ $D3$-branes. At low energies, the dynamics on the $D$-branes
reduces to $N=4$ supersymmetric Yang -- Mills theory with gauge group $U(N)$. At 
weak 't Hooft coupling $\lambda$, one can study it using standard tools of perturbative 
field theory, while for large $\lambda$ it is well described by IIB supergravity (or
the corresponding string theory) on $AdS_5\times S^5$ with a large radius of curvature 
$R$. This geometry plays the role of our bulk geometry (\eg\ \spmet) for this case. 

Consider further a $D$-string stretched in a direction orthogonal to the threebranes, 
ending on them at a point in $\IR^3$. From the perspective of the $U(N)$ gauge theory 
on the $D3$-branes, this string describes a magnetic monopole, charged under a $U(1)$
in $U(N)$. It preserves half of the supersymmetry of the theory, and can be thought
of as an analog of the BPS branes in our discussion. Like those branes, it is stretched
in the radial (holographic) direction. 

At weak coupling, one can describe this monopole as a source of magnetic field, and one
of the scalar fields on the threebranes \CalMal. Introducing a localized perturbation of one
of the transverse scalar fields on the $D$-string corresponds in the gauge 
theory to a change in the electromagnetic field around it, which can be induced for example
by wiggling the monopole. Such perturbations can be analyzed directly in gauge theory; they 
typically propagate away from the monopole as time goes by.  

Note that both the monopole and the perturbations around it have energy that goes like 
$1/g_{YM}^2$; this is the analog of the fact that in our discussion in the previous sections,
we studied excitations with energy of order $1/g_s$. Here this is simply the statement
that they are classical gauge theory effects.

The direct analog of the discussion of the previous sections in this case is the $D1-D3$ 
system at strong coupling (\ie\ large $\lambda$). Now the monopole corresponds to a 
$D$-string stretched in the radial direction of $AdS_5$. Localized excitations of the
$D$-string can be studied using its DBI action. Parameterizing the Poincare 
patch of $AdS_5$ by the boundary coordinates $(x^0,x^1,x^2,x^3)$ and radial coordinate $z$, 
and considering $z$-dependent fluctuations of the position of the string in the $x^1$ direction, 
this action takes the form  
\eqn\NGAct{
S=-\tau^{BPS}_1R^2 \int \frac{dtdz}{z^2}\sqrt{1-{\dot x}_1^2+(x_1')^2}~.
}
Here dot and prime denote derivatives w.r.t. $t$ and $z$, respectively.
 
The action \NGAct\ is an analog of \ExpAct\ for this case, and we can proceed as in section 4. 
Expanding \NGAct\ to quadratic order in ${x}_1$ leads to the Klein-Gordon equation 
\eqn\kkggee{-\ddot x_1+ x_1''-\frac{2}{z} x_1'=0~,}
which is an analog of \LinearIn\ for this case. The general solution of \kkggee\ has the form \CalGui
\eqn\CalGuSoln{x_1=f(z-t)-zf'(z-t) ~,}
with $f$ an arbitrary function of its argument. We see that small perturbations with finite support 
move to large $z$ (\ie\ away from the boundary of $AdS_5$), while growing in the process. Eventually,
the linear approximation \kkggee\ breaks down, and one has to go back to the full non-linear EL equation
of \NGAct, like in our discussion in section 4. 

The solution \CalGuSoln\ is completely determined by the behavior of $x_1$ at the boundary $z=0$, 
$x_1(z=0,t)=f(-t)$. Thus the dynamics of the $D$-string can be recovered from the trajectory of the 
monopole. Note also that in this case it is clear that the perturbation \CalGuSoln, which grows with $z$,
corresponds to the time-dependent gauge field around the moving monopole. This is a manifestation of the 
open string completeness proposal -- the time-dependent profile of open string fields on a $D$-brane 
describes the evolution of closed strings with energy of order $1/g_s$ (the gauge fields in a strongly 
coupled gauge theory). 

Brane profiles with growing $x_1(t,z)$ have also been encountered in  computations of  the drag force on 
a moving  quark \refs{\HKKKY,\Drag}. In that case $x_1(t,0)$ grows as well (while in ours $x_1(t,0)$ always
remains small), but the qualitative interpretation of the ``jets" of energy moving away from the monopole
\CalGuSoln\ should be the same. The moving monopole produces a directed flow of the plasma; the growth
of $x_1(t,z)$ should be interpreted as propagation of this flow away from the source. Since this growth occurs at large values of $z$, the flow also expands in the $x_2$ and $x_3$ directions.\foot{The authors of \HKKKY\ use a nice analogy with waves behind a moving boat.} 

The $D-\bar D$ system discussed in section 5 corresponds in the $D1-D3$ system to a monopole -- antimonopole 
pair. The system of a $D$-string and antistring separated by a distance $L$ has a higher energy than that of 
a single $D$-string connecting the monopole and antimonopole for all $L$ \MaldRey. Thus, the reconnection 
discussed in section 5 always occurs. This is natural since the $D3$-brane theory has no scale.

\newsec{Discussion}

The main point of this paper is that the tachyon Dirac-Born-Infeld action
\tdbiA\ provides a useful description of BPS and non-BPS $D$-brane dynamics
in critical string theory. It can be thought of as describing branes living
in a $10+1$ dimensional spacetime. BPS branes are extended in the extra
dimension, labeled by $T$, while non-BPS ones are localized in it.

The action \tdbiA\ reproduces many aspects of brane dynamics. For non-BPS
branes these include \SenNF\ the mass of the open string tachyon, some
properties of the static classical solution \soleuc, which is obtained by
condensing the open string tachyon on the brane, and the rolling tachyon
solution  \rolling. For BPS $D$-branes it reproduces the tension of
the brane, \taubps, the DBI action \ssdbi\ describing the dynamics of the
light modes, and their coupling to light closed string modes.

The TDBI action is particularly useful for studying the $D-\bar D$
system. It incorporates the $D-\bar D$ tachyon, and describes correctly
many of its properties. For example, it gives the correct value of the
brane -- antibrane separation, $L$, for which the tachyon becomes massless,
\llccrr, and captures the interactions between the tachyon
and the light fields on the brane and antibrane. It also gives a nice geometric
picture of $D-\bar D$ tachyon condensation, as a process in which the brane
and antibrane connect at large $|T|$, as in figs. 3, 4, 7.

The description of $D$-brane dynamics in terms of the TDBI action also sheds light
on the open string completeness proposal of A. Sen \SenNF. It suggests that
this proposal should be thought of as a generalization of open-closed string
duality to energies of order $1/g_s$. At those energies, the dynamics of closed
strings (\ie\ states that are not confined to a $D$-brane) should have an open
string description; the TDBI action provides such a description for states that
couple to the $D$-brane.

Some elements of the picture presented in this paper require further work.
An important question is the relation between the TDBI description of the tachyon,
as a geometric mode  associated with deformations of the $D-\bar D$ system  of the
sort depicted in \DDbartach, and the standard description in terms of a complex
field $\CT$ with mass \masstt. We proposed that the usual description is useful
for small values of the tachyon, $\CT\ll1$, while the TDBI one is more useful for
$\CT\gg 1$, a regime in which the standard description receives large $\alpha'$
corrections. Thus, the relation between the two descriptions is a kind of worldsheet
duality.

It is interesting to understand the relation between the tachyon field $\CT$ and the
massless field $\phi$, \Avars, which describes the $D-\bar D$ tachyon for $L=L_{\rm cr}$.
By comparing the charges of $\CT$ and $\phi$ under the $U(1)$ gauge fields $A_\mu^{(L)}$,
$A_\mu^{(R)}$, we proposed that $\CT\propto\phi^2$, \reltachphi. We would next like to
offer another argument that supports this identification and generalizes it to other values of $L$.

In section 5 we saw that for $L<L_{\rm cr}$ $(L>L_{\rm cr})$ the lowest lying mode
associated with the reconnection process of \DDbartach\ is tachyonic (massive). Its mass
can in principle be computed by studying generalizations of the time-dependent solution
\timedepsol\ to other energies. We have not constructed such solutions, but suppose this
was done, and a generalization of the field $\phi$ \Avars\ was identified. If this field
is described by a massive Klein-Gordon Lagrangian, one can deduce its scaling with $T$
by comparing to the TDBI Lagrangian \KofAct.

This implies that $\phi$ must scale like\foot{Note that this is indeed the case for \Avars,
using \soleuc.}
\eqn\scalephi{\phi\sim e^{-{\alpha T\over2}}}
as $T\to\infty$. Using \reltachphi\ we conclude that the open string tachyon $\CT$ scales like
$\exp(-\alpha T)$. The mass of $\CT$ can then be read off \timedepsol, by matching the time
dependence of the solution to that expected for a Klein-Gordon field. This gives
\eqn\masssttt{m_T^2=-a^2\alpha^2=-\alpha^2+{\alpha^4L^2\over\pi^2}~.}
For $\alpha=1/\sqrt2$, the case relevant for critical string theory, \masssttt\ agrees
with the exact string theory result  \masstt. This provides independent evidence for the
identification \reltachphi. To further substantiate it, one would need to understand
better the time-dependent solutions for the configurations of \DDbartach.

More generally, it would be interesting to understand the nature of the extra dimension, labeled
by $T$, which plays an important role in our construction. As we discussed, this dimension
seems to play a similar role to that of the holographic direction in gauge-gravity duality.
Its emergence seems to be related to the excess energy above the supersymmetric vacuum
contained in the non-BPS $D$-brane and $D-\bar D$ systems that we study. 

It would also be interesting to extend the discussion to closed strings. There are many
known vacua of string theory which have closed string tachyons, \eg\ the type 0 theories
described in \PolchinskiRR. It is natural to ask whether in this case too, the tachyon
field parameterizes an extra dimension associated with energy above the supersymmetric
(type II) vacuum.

The idea that adding energy to supersymmetric vacua may lead to the appearance of additional
spatial dimensions is reminiscent of \HellermanZM, although there is no obvious relation between
that work and ours.

As mentioned in the introduction, $D-\bar D$ systems play an important role in many applications.
An example is the Sakai-Sugimoto model of holographic QCD \refs{\SakaiCN,\AntonyanVW}, which
describes massless chiral fermions with strong interactions between the left and right handed
fermions, that resemble those of QCD and the Nambu-Jona-Lasinio model in different regions of
its parameter space. An outstanding problem is to add mass to these fermions. This involves turning
on an expectation value for the $D-\bar D$ tachyon on a $D8-\bar D8$ brane pair (see \eg\
\refs{\AharonyAN,\McNeesKM} for some recent discussions).

In the usual treatment of the Sakai-Sugimoto model, the eightbranes are described by their
DBI action in a curved background. Since, as we showed in section 3, the TDBI action reduces
to the DBI one for branes extended in $T$, all the successes of that treatment carry over
to our construction. However, since our model includes the $D-\bar D$ tachyon, it should be
possible in it to study the quark mass deformation.

Another application where our construction may be useful is brane-antibrane inflation.
In that context, the stage of the dynamics in which the $D-\bar D$ tachyon becomes light
and condenses is much less understood than the stage in which the branes approach each
other due to their gravitational attraction. Our construction provides a field theoretic
model which can be used to analyze the dynamics of this stage in more detail.

Another aspect of our work that might be interesting for inflation is the $D-\bar D$ system
in the background of $k$ $NS5$-branes described above. In this case, the energy scale associated
with the dynamics is reduced by a factor of $\sqrt k$. For example, the distance at which
the $D-\bar D$ tachyon becomes tachyonic is given by $L_{\rm cr}=\pi\sqrt{k} l_s$, \largel.
This might be useful for lowering the scale of inflation.

Systems of $D$-branes and antibranes ending on $NS5$-branes are ubiquitous in brane
constructions of non-supersymmetric vacua of SQCD \refs{\OoguriBG-\GiveonEW}. The process
of reconnection of the branes discussed in section 5 happens in such systems as the parameters
of the brane configuration are varied. Our results can be used to study such processes in more
detail.

\vskip 1cm
\centerline{\bf Acknowledgements}

\noindent
This work was supported in part by DOE grant DE-FG02-90ER40560, by the
National Science Foundation under Grant 0529954, and by the Israel-U.S.
Binational Science Foundation. The work of DE was supported in part by a 
GAANN fellowship from the Department of Education. DK thanks the Weizmann 
Institute for hospitality during part of this work.

\appendix{A}{Analysis of solutions of \actyy}

In this appendix we study the TDBI action for a single translational mode
on a BPS $D$-brane, \actyy. We start with a discussion of the linear regime,
where one can approximate this action by \quadrAct, and then move on to the
full non-linear problem.

\subsec{No normalizable modes}

As discussed in section 4, for a mode with mass $m$, the $T$-dependent wavefunction
$z(T)$ \sepvar\ satisfies the Klein-Gordon equation \phimm.
Finiteness of the energy density,
\eqn\quadrAct{
{\cal E}=\frac{1}{2}\int dTV(T)\left[(\partial_t y)^2+(\partial_i y)^2+(\partial_T y)^2\right]<\infty~,
}
requires the finiteness of
\eqn\normalCond{
\int dTV(T)(z')^2,\quad
\int dTV(T)z^2~.
}
Substituting the form of the potential \genv, rewriting \phimm\ in terms of the variable
\eqn\defxx{x=\sinh(\alpha T)~,}
and imposing the finiteness of \normalCond\ leads to the eigenvalue
problem:
\eqn\eigenEqn{
(1+x^2)z''+\frac{m^2}{\alpha^2} z=0,\quad
\int_{-\infty}^\infty \frac{dx}{1+x^2}z^2<\infty~.
}
The general solution of \eigenEqn\ takes the form:
\eqn\aaaa{
z=(1+x^2)^{\frac{1}{4}+s}\left[
c_1F(-\frac{1}{4}-s,\frac{3}{4}-s;\frac{1}{2};\frac{x^2}{x^2+1})+
\frac{c_2 x}{\sqrt{1+x^2}}
F(\frac{1}{4}-s,\frac{5}{4}-s;\frac{3}{2};\frac{x^2}{x^2+1})\right]~,
}
where $F$ is the hypergeometric function, $c_1$, $c_2$ are arbitrary constants, and
\eqn\ssss{
s=\frac{1}{4}\sqrt{1-\frac{4m^2}{\alpha^2}}~.
}
For the integral in \eigenEqn\ to converge as $x\to\pm\infty$, one must have
\eqn\finint{
F(-\frac{1}{4}-s,\frac{3}{4}-s;\frac{1}{2};1)=0:\qquad
s=\frac{1}{4}-n,\quad n\in Z_+~.
}
Since $s$ defined by \ssss\ is non-negative, we conclude that the only normalizable mode is the one with
$n=0$, $s=\frac{1}{4}$, which corresponds, according to \ssss, to a massless mode with a constant wavefunction
$z(T)=$const. This is one of the modes discussed in section 3.

Note that for $m^2>\alpha^2/4$ the parameter $s$ \ssss\ becomes imaginary, and for large $x$
one has
\eqn\contin{
z_m\approx c\sqrt{x}e^{2s\ln x}~.
}
The corresponding wavefunctions are delta-function normalizable. They give rise to the
continuum above a gap that is familiar from linear dilaton backgrounds (to which our
problem is closely related, as explained in the text).

\subsec{Solutions of the massive Klein -- Gordon equation}

In this subsection we will construct a general solution of \MassKG\ and discuss some of its 
properties. We will be interested in configurations which initially have zero velocity (\ie, 
${\dot w}(0,T)=0$). The general solution of \MassKG\ with such initial condition has the form
\eqn\PreSolveKG{
w(t,T)=\int dk \phi(k)e^{ikT}\cos\omega t=\frac{1}{2\pi}
\int dy w(0,y)\int dk e^{ik(T-y)}\cos\omega t~.
}
Here we introduced $\phi(k)$, the Fourier  transform of the initial waveform $w(0,T)$.  
The frequency $\omega$ is defined by
\eqn\KGOmega{\omega^2=k^2+m^2~.}
To simplify  \PreSolveKG, we first assume that the point $(t,T-y)$ belongs to the forward light 
cone. In this case, it is convenient to introduce the variables $\gamma$, $\beta$, $\lam$:
$$
k=m~\sinh\gamma,\quad \lam^2=t^2-z^2>0,\quad
T-y=\lam~\sinh\beta~,
$$
and to use the standard integrals
\eqn\Hankel{
\eqalign{
&\frac{1}{m}\int dk e^{ikz+i\omega t}=\int e^{im\lam\cosh\gamma}
\cosh(\gamma-\beta)
d\gamma=-\frac{\pi}{2}\cosh\beta~H^{(1)}_1(m\lam)~,
\cr
&\frac{1}{m}\int dk e^{ikz-i\omega t}=\int e^{-im\lam\cosh\gamma}
\cosh(\gamma+\beta)
d\gamma=-\frac{\pi}{2}\cosh\beta~H^{(2)}_1(m\lam)~.
}}
Here $H^{(1)}_1(z)$ and $H^{(2)}_1(z)$ are Hankel functions. Substituting the integrals 
\Hankel\ into \PreSolveKG, one finds that in the interior of the light cone $w$  takes the
form
\eqn\SemiSolveKG{t^2-(T-y)^2>0:\qquad
w(t,T)=-\frac{mt}{4}\int dy w(0,y)\frac{J_1(m\lam)}{\lam}~,
}
where 
$$
J_1(z)=\frac{1}{2}\left[H^{(1)}_1(z)+H^{(2)}_1(z)\right]
$$
is a Bessel function. 

If $(t,T-y)$ is a space--like vector, then the parameter $\beta$ should be introduced differently 
($T-y=\lam\cosh \beta$, $t=\lam\sinh\beta$), and, in accordance with expectations, we conclude 
that the right--hand side of \PreSolveKG\ vanishes,
\eqn\ZeroHank{
\eqalign{
&\int dk e^{ikz+i\omega t}=m\int e^{im\lam\sinh\gamma}
\cosh(\gamma-\beta)d\gamma~,\cr
&\int dk e^{ikz-i\omega t}=m\int e^{-im\lam\sinh\gamma}
\cosh(\gamma+\beta)d\gamma=-\int dk e^{ikz+i\omega t}~.
}}
To evaluate \PreSolveKG\ on the light cone (where $T-y=t$),
we differentiate that expression with respect to $m$:
\eqn\HankLC{
\eqalign{
&\d_m\int dk e^{ikt+i\omega t}=
mt\int \frac{idk}{\sqrt{k^2+m^2}} e^{ikt+i\omega t}=
imt\int_0^\infty e^{iut}\frac{du}{u}~,
\cr
&\d_m\int dk e^{ikt-i\omega t}=
-mt\int \frac{idk}{\sqrt{k^2+m^2}} e^{ikt-i\omega t}=
-imt\int_0^\infty e^{-ivt}\frac{dv}{v}~,
\cr
&\left.\d_m\int dk e^{ikx}\cos(\omega t)\right|_{x=t}=
-mt\frac{\pi}{2}~.
}}
Combining this with the result for $m=0$ (which gives a $\delta$-function),
we find the value of the integral at $T-y=t$:
$$
\left.\int dk e^{ikx}\cos(\omega t)\right|_{x=t}=\pi\delta(x-t)-
\frac{m^2t}{4}\pi~.
$$
Together with \SemiSolveKG, this equation gives the most general solution of the Klein--Gordon 
equation \MassKG\ with $\dot w(0,T)=0$:
\eqn\SolveKGMain
{
w(t,T)=\frac{1}{2}w(0,T-t)+\frac{1}{2}w(0,T+t)-
\frac{mt}{4}\int_{|T-y|<t} dy w(0,y)\frac{J_1(m\lam)}{\lam}~,
}
where
\eqn\deflam{
\lam^2=t^2-(T-y)^2~.
}
For any initial waveform with finite support, this solution has the following properties\foot{We concentrate 
only on the part of the waveform which moves towards positive $T$. A second part moves in the opposite 
direction, but it will not be important for our discussion.}:
\item{1.}
At late times,  $w(t,T)$ goes to zero
at any given $T$. Indeed, assuming that $w(0,y)$ vanishes at
$|y|>y_0$, one can replace $\lam$ by $t$ at times $t\gg |T+y_0|$. This
simplifies the integral in \SolveKGMain:
$$
w(t,T)\sim -\frac{m}{4}J_1(mt)\int dy w(0,y)\sim
-\frac{m}{2\sqrt{\pi mt}}\sin (mt)\int dy w(0,y)
$$
and demonstrates that $w(t,T)$ goes to zero at late times.

\item{2.} For $T\sim t$, the function $w$ oscillates, and its periods in $T$ become smaller as 
time goes by. To show this, we look at the extrema of  $w$. Their locations, $T(t)$, 
obey an integral equation which is obtained by differentiating \SolveKGMain\ with respect to $T$:
\eqn\MotOfMax{
T=T(t):\quad \frac{1}{2}w'(0,T-t)-
\frac{mt}{4}\d_T \int dy w(0,y)\frac{J_1(m\lam)}{\lam}\theta(\lam^2)=0~.
}
To extract qualitative properties of $T(t)$, we take $w(0,x)$ to be
a rectangle of width $2a$ (\ie\ $w(0,x)=\theta(x+a)-\theta(x-a)$,
see  \linFig). This allows us to evaluate the integral
in the last equation. The waveform has two sharp edges at $T=t\pm a$
and a sequence of additional extrema $T(t)$. For $T\ne t\pm a$,
equation \MotOfMax\ can be rewritten as \eqn\MaxBessel{
\frac{J_1(m\lam_+)}{\lam_+}\theta(\lam_+^2)-
\frac{J_1(m\lam_-)}{\lam_-}\theta(\lam_-^2)=0,\quad
\lam_{\pm}=\sqrt{t^2-(T\pm a)^2}~. } 
Since the function $\frac{J_1(m\lam)}{\lam}$ goes to zero at large $\lam$,
the relation
$$
\lam_-^2=\lam_+^2+2Ta
$$
implies that the supports of the two terms in \MaxBessel\ become
well-separated at late times ($\lam_-$ is large when $\lam_+$ is small and vice versa). 
Thus, local extrema of $w$ correspond to $J_1(m\lam_+)=0$ and to $J_1(m\lam_-)=0$.
This gives the positions of the peaks in terms of zeroes of the Bessel function:
$$
T_n(t)=\pm a+\sqrt{t^2-\frac{c_n^2}{m^2}}\approx
\pm a+t-\frac{c^2_n}{2m^2t},\quad J_1(c_n)=0~.
$$
As time progresses, all peaks approach $T_0=\pm a+t$, so the pulses are getting narrower. 
The velocities of the peaks in the $T$ direction, ${\dot T}_n(t)$, approach the speed of light.

\item{3.} To determine the height of the peaks, we again assume that $w(0,x)$ is a combination 
of step-functions ($w(0,x)=\theta(x+a)-\theta(x-a)$), and evaluate \SolveKGMain\ at the locations 
of the peaks:
\eqn\IntegrJOne{
w(t,T_n(t))=-\frac{mt}{4}\int_{-a}^a dy \frac{J_1(m\lam_n)}{\lam_n}
\theta(t-|T_n(t)-y|)~.
}
To simplify this expression, we make an approximation for
$\lam$ at $T=T_n(t)$:
$$
\lam_n\equiv\left[t^2-(T_n(t)-y)^2\right]^{1/2}\approx
\left[t^2-(t+a-\frac{c_n}{2m^2 t}-y)^2\right]^{1/2}\approx
\sqrt{2t(y-a+\frac{c_n}{2m^2 t})}
$$
and define a new variable $z\equiv 2t(y-a+\frac{c_n}{2m^2 t})$. The contribution to the integral 
in \IntegrJOne\ comes from the region where $\lam_n^2>0$:
$$
a>y>a-\frac{c_n}{2m^2 t}\quad\rightarrow\quad
\frac{c_n}{m^2}>z>0~.
$$
Rewriting \IntegrJOne\ in terms of the variable $z$, we find the late time behavior of the peaks:
\eqn\IntegrJTwo{
\eqalign{
w(t,T_n(t))\sim
-\frac{m}{8}\int_0^{c_n^2/m^2} dz
\frac{J_1(m\sqrt{z})}{\sqrt{z}}~.
}}
This calculation demonstrates that the peaks of $w$ approach some finite height, leading to exponential growth of the peaks of $y$  (see \rescyy, \maxyy). In particular, any initial perturbation approaches a 
point where the derivatives of $y$ become large and the linear approximation \LinearIn\ breaks down.

\appendix{B}{Interaction between the $D-\bar D$ tachyon and the translational mode}

We saw in section 5 that the massless scalar field corresponding to the distance between 
a $D$-brane and a $\bar D$-brane on a circle becomes massive when a non-zero tachyon
vev is turned on (see discussion at the end of subsection 5.1). In this appendix we determine 
its mass in the TDBI approximation. To do this, we consider small fluctuations around the profile \soleuc:
\eqn\GenerSLuc{
\sinh\alpha T=A\cos \alpha y+f(x^\mu,y)~.
}
Substituting this into \actyy\ and expanding  to quadratic order in $f$, we arrive at the action 
for $f(x^\mu,y)$:\foot{The easiest way to derive this action is to rewrite \GenerSLuc\ as 
$\sinh\alpha T=A\cos[\alpha (y-g(x^\mu,T))]$, find the quadratic action for $g$, and use the 
relation $f=Ag\alpha\sin\alpha y$.}
\eqn\TranlAct{
S=-\frac{\tau_p}{2\alpha^2}\int d^{p+1}x dy
\left\{\frac{\sin^2\alpha y}{(1+A^2)^{3/2}}\left[\d_y\frac{f}{\sin\alpha y}\right]^2+
\frac{\d_\mu f\d^\mu f}{
\sqrt{1+A^2}(1+(A\cos \alpha y)^2)}\right\}~.
}
Separating variables ($f(x^\mu,y)={\hat f}(x^\mu)\phi(y)$) 
and defining the $p+1$ dimensional mass $m$  ($\d_\mu \d^\mu {\hat f}=m^2{\hat f}$), 
the equation of motion for \TranlAct\ reduces to
\eqn\RDOne{
\d_y^2 \phi+\alpha^2\phi+\frac{m^2(1+A^2)}{1+(A\cos \alpha y)^2}\phi=0~. 
}
We are interested in solutions which are periodic in $y$: 
$\phi(y+\frac{2\pi}{\alpha})=\phi(y)$. 

Equation \RDOne\ has two solutions that are massless for all values of $A$,
\eqn\RDNoMass{
\phi_1(y)=\cos\alpha y~,\qquad \phi_2(y)=\sin\alpha y~,
}
corresponding to shifts of $A$ and  $y$, respectively. It also describes a sequence of 
massive modes. One of these modes becomes light as $A\to\infty$, where it corresponds 
to the distance between the brane and antibrane. Our purpose here is to determine the 
mass of this mode for large but finite $A$.

Near the locations of the branes the light mode is expected  to behave as:
\eqn\LightEnds{\phi\left(y\sim \pm {\pi\over2\alpha}\right)\sim
\pm\sin(\alpha y)}
In particular, one has
$\phi'(\frac{\pi}{2\alpha})=\phi'(-\frac{\pi}{2\alpha})=0$ and 
$\phi(\frac{\pi}{2\alpha})=\phi(-\frac{\pi}{2\alpha})$. Since equation 
\RDOne\ and the boundary conditions  \LightEnds\ are invariant under 
$y\to -y$, $\phi$ is an even function of $y$, and $\phi'(0)=0$. 

To make the formulas appearing below more compact, it is convenient to introduce 
a shifted variable $z=\frac{\pi}{2}-\alpha y$, in terms of which \RDOne\ takes the form
\eqn\RDOnePr{
\d_z^2 \phi+\phi+\frac{m^2(1+A^2)}{\alpha^2(1+(A\sin z)^2)}\phi=0~,
\qquad \phi'(0)=\phi'(\frac{\pi}{2})=0~. }
Assuming that $1\gg m\gg 1/A$, this equation can be simplified in two overlapping regions:
\eqn\RDTwo{\eqalign{
(A\sin z)^2\ll (mA)^{2}~:&\qquad
\d_z^2 \phi_1+\frac{m^2 A^2}{\alpha^2(1+(Az)^2)}\phi_1
=0~, \cr
(A\sin z)^2\gg 1~:&\qquad
\d_z^2 \phi_2+\phi_2+\frac{m^2\phi_2}{\sin^2 z}=0~.
}}
The solution satisfying the boundary condition $\phi_1'(0)=0$ can be expressed in terms of 
the hypergeometric function:
\eqn\RDThree{\eqalign{
&\phi_1(z)=F(-\frac{1+s}{4},-\frac{1-s}{4};\frac{1}{2};
-(Az)^2)~,\quad 
s=\sqrt{1-\frac{4m^2}{\alpha^2}}~.
}}
To match this with $\phi_2(z)$, we need to evaluate $\phi_1(z)$ at 
$Az\gg 1$. Using formulae for analytic continuation of the hypergeometric function, we find:
\eqn\RDThreeA{\eqalign{
\phi_1(z)&=(1+A^2 z^2)^{\frac{1+s}{4}}F\left[
-\frac{1+s}{4},\frac{3-s}{4};\frac{1}{2};
1-\frac{1}{1+A^2z^2}\right]\cr
&=(A z)^{\frac{1+s}{2}}\left\{
\frac{\Gamma(\frac{1}{2})\Gamma(\frac{s}{2})}{\Gamma(\frac{s+3}{4})\Gamma(\frac{s-1}{4})}\left[1+O(\frac{1}{A^2 z^2})\right]+
\frac{\Gamma(\frac{1}{2})\Gamma(-\frac{s}{2})}{\Gamma(\frac{3-s}{4})\Gamma(-\frac{1+s}{4})}(Az)^{-s}\left[
1+O(\frac{1}{A^2 z^2})
\right]\right\}\cr
&\approx \frac{(s-1)\pi}{4}(A z)^{\frac{1+s}{2}}+(A z)^{\frac{1-s}{2}}~.
}}
At the last step we used the fact that $s$ is close to one and kept only the leading terms in the coefficients. 

Turning to $\phi_2$, the general solution is again expressed in terms of hypergeometric functions:
\eqn\RDFour{\eqalign{
\phi_2(z)=&c_1\sin^{\frac{1+s}{2}}z
F\left(\frac{s-1}{4},\frac{3+s}{4};1+\frac{s}{2};\sin^2 z\right)\cr
+&c_2\sin^{\frac{1-s}{2}}z
F\left(\frac{-s-1}{4},\frac{3-s}{4};1-\frac{s}{2};\sin^2 z\right)~.
}}
Comparing to \RDThreeA\ leads to a relation between 
$c_1$ and $c_2$:
\eqn\RDThreeCC{
c_2=\frac{4}{\pi(s-1) A^{s}}c_1\approx \frac{4}{\pi(s-1) A}c_1.
}
The value of $s$ is determined by imposing the boundary condition 
$\phi'_2(\frac{\pi}{2})=0$. Indeed, using
\eqn\RDSix{
\sin^{\frac{1+s}{2}}z
F\left(\frac{s-1}{4},\frac{3+s}{4};1+\frac{s}{2};\sin^2 z\right)={\rm const}+
\frac{\Gamma(\frac{2+s}{2})\Gamma(-\frac{1}{2})}{
\Gamma(\frac{s-1}{4})\Gamma(\frac{3+s}{4})}\cos z+O(\cos^2 z)
}
and its counterpart with $s$ replaced by $-s$, we conclude that 
vanishing of $\phi'_2(\frac{\pi}{2})$ is equivalent to the relation
\eqn\RDSev{
c_2=-\frac{\Gamma(\frac{2+s}{2})\Gamma(-\frac{1}{2})}{
\Gamma(\frac{s-1}{4})\Gamma(\frac{3+s}{4})}~
\frac{
\Gamma(\frac{-s-1}{4})\Gamma(\frac{3-s}{4})}{\Gamma(\frac{2-s}{2})\Gamma(-\frac{1}{2})}~c_1\approx \frac{\pi(s-1)}{4}c_1
}
Comparing this with \RDThreeCC, we find an expression for 
$s$:
\eqn\RDEgt{
s=1-\frac{4}{\pi\sqrt{A}}~.
}
Using \RDThree\ we find the mass of the relative translational mode
\eqn\RDNine{
m^2=\frac{2\alpha^2}{\pi\sqrt{A}}={\alpha^3\phi\over\pi}\left(2\tau_{p-1}^{BPS}\right)^{-\half}~.
}

\listrefs
\end